\documentclass[aps,showpacs,amsmath,amssymb,eqsecnum,twocolumn]{revtex4}

\usepackage{bm}
\usepackage{graphicx}
\begin{document}
\title{Casimir force in $O(n)$ lattice models with a diffuse interface}

\author{Daniel Dantchev}
\affiliation{%
Fachbereich Physik, Universit{\"a}t Duisburg-Essen, Campus Duisburg, D-47048 Duisburg,
Germany
}%
\affiliation{%
Institute of Mechanics---BAS, Acad.\ G.\ Bonchev St.\
bl.~4, 1113 Sofia, Bulgaria}
\author{Daniel Gr{\"u}neberg}
\affiliation{%
Fachbereich Physik, Universit{\"a}t Duisburg-Essen, Campus Duisburg, D-47048 Duisburg,
Germany
}%

\date{\today}

\begin{abstract}
On the example of the spherical model we study, as a function of the temperature $T$, the behavior of the Casimir force in $O(n)$ systems with a diffuse interface and slab geometry $\infty^{d-1}\times L$, where $2<d<4$ is the dimensionality of the system. We consider a system with nearest-neighbor anisotropic   interaction constants $J_\parallel$ parallel to the film and $J_\perp$ across it. The model represents the $n\to\infty$ limit of $O(n)$ models with antiperiodic boundary conditions applied across the finite dimension $L$ of the film. We observe that the Casimir amplitude $\Delta_{\rm Casimir}(d|J_\perp,J_\parallel)$ of the anisotropic $d$-dimensional system is related to that one of the isotropic system $\Delta_{\rm Casimir}(d)$ via $\Delta_{\rm Casimir}(d|J_\perp,J_\parallel)=\left(J_\perp/J_\parallel\right)^{(d-1)/2} \Delta_{\rm Casimir}(d)$. For $d=3$ we find the exact Casimir amplitude
$ \Delta_{\rm Casimir}= \left[ {\rm Cl}_2\left (\pi/3 \right)/3-\zeta
   (3)/(6 \pi )\right]\left(J_\perp/J_\parallel\right)$,
as well as the exact scaling functions of the Casimir force and of the helicity modulus $\Upsilon(T,L)$. We obtain that $\beta_c\Upsilon(T_c,L)=(2/\pi^{2}) \left[{\rm Cl}_2\left(\pi/3\right)/3+7\zeta(3)/(30\pi)\right]
\left(J_\perp/J_\parallel\right)L^{-1}$, where $T_c$ is the critical
temperature of the bulk system. We find that the effect of the helicity is thus strong that the Casimir force is repulsive in the whole temperature region.
\end{abstract}
 \maketitle

\section{INTRODUCTION}

The excess free energy due to the finite-size contributions to the free
energy of a system with a film geometry characterizes a fluctuation-mediated
interaction which is termed the Casimir force, or, in the case of a fluid
confined between two parallel walls - also the solvation force (or the disjoining
pressure). The force is named so after the Dutch physicist Hendrik B. G. Casimir who in 1948 \cite{C48} first noticed that when two metallic perfectly conducting uncharged plates face each other in
vacuum at zero temperature the restriction and the modification of the
zero-point  vacuum fluctuations of the electromagnetic field between
the two parallel  plates  lead   to a dependence of the energy of
the system on the distance $L$ between the plates and, thus, to a
force between them which turns out to be attractive. The above is the so-called classical (actually
quantum mechanical) Casimir effect. When the fluctuating medium is not a vacuum, but a thermodynamic system, say fluid, near its bulk critical point $T_c$ one arrives at the so-called thermodynamic Casimir effect that has been predicted by M. E. Fisher and P. G. de Gennes \cite{FdG78} in 1978  and which has been a subject of intensive theoretical and experimental studies  afterwards \cite{evans,K94,BDT2000,KD92,D98,M97,M99,GC,GCmixture,ML,UBMCR03, DKD2003,SHD03,DK2004,FYP05,DDG2006,DGS2006,GSGC2006,DSD2007,VGMD2007,H2007,GD2008,HHGDB2008}.

We remind that for an $O(n),n\geq 1$ model of a $d$-dimensional system with a temperature $T$ and geometry $\infty^{d-1}\times L$ the Casimir force is defined by  \cite{evans}, \cite{BDT2000}
\begin{equation}
F_{\rm Casimir}^{(\tau )}(T,L)=-\frac{\partial f_{\rm ex}^{(\tau )}(T,L)}{\partial L}%
\text{,}  \label{def}
\end{equation}
where $f_{\rm ex}^{(\tau)}(T,L)$ is the excess free energy
\begin{equation}
f_{\rm ex}^{(\tau )}(T,L)=f^{(\tau)}(T,L)-Lf_b(T)\text{,}  \label{fexd}
\end{equation}
and the superscript $\tau $ denotes the dependence on the boundary
conditions. Here $f^{(\tau )}(T,L)$ is the full free energy per unit area of such a system under boundary conditions $\tau $ and $f_b
$ is the bulk free energy density. It is believed that if the boundary
conditions $\tau $ are the same at the both bounding the system surfaces $%
F_{\rm Casimir}^{(\tau )}$ will be negative. In the case of a fluid confined between identical walls this implies that then the net force between the plates will be {\em attractive} for large separations. If the boundary conditions  are essentially different  at the both confining the system surface planes (e.g. one of the surfaces prefer the liquid phase of the fluid while the other prefers the gas phase) the Casimir force is expected to be positive in the whole region of the thermodynamic parameters, i.e. then the net force between the plates will be {\it repulsive}.

In the current article we will investigate the behavior of the Casimir fore in systems with diffuse interface. As a realization of such systems one can think of about the reaction of $O(n)$ models with $n\ge 2$ to some helical external field which reaction can be characterized in terms of some helicity modulus $\Upsilon$ or, in case of a magnetic materials, of Bloch walls between the domains of the magnet. Heuristically, the helicity modulus is the analog of the interface
tension for $O(n)$-symmetric systems. The simplest theoretical model of a system with a diffuse interface is the $O(n)$ model with antiperiodic, i.e. $\tau \equiv a$, boundary conditions and short-ranged interactions. According to the standard finite-size scaling theory (see, e.g., \cite{BDT2000,privman} for a general review) one expects that near the critical temperature $T_c$ (of the corresponding bulk, i.e. $L=\infty$ system) the behavior of $F_{\rm Casimir}^{\left( a\right) }$ will be given by
\begin{equation}
\beta F_{\rm Casimir}^{\left(a\right) }(T,L)=L^{-d}X_{\rm Casimir}^{\left(a\right)
}(x_t)\text{,}  \label{fssc}
\end{equation}
while that one of the full free energy $f^{(a)}$ is
\begin{equation}\label{fssfe}
\beta f^{(a)}(T,L)=L^{-(d-1)} X_{f}^{\left(a\right)}(x_t)\text{,}
\end{equation}
where $x_t=a_t tL^{1/\nu }$ is the temperature scaling variable with $t=(T-T_c)/T_c$ being the reduced temperature, $a_t$ is a nonuniversal scaling factor, while $X_{\rm Casimir}^{\left( a\right) }$ and $X_{f}^{\left(a\right)}$ are {\em universal} (geometry
dependent) scaling function and $\nu $ is corresponding
(universal) scaling exponent that characterizes the temperature divergence of the correlation length $\xi_b$ when one approaches the bulk critical temperature from above, i.e. $\xi _b(t\rightarrow 0^+)\simeq \xi_0^+ t^{-\nu}$. The scaling functions $X_{\rm Casimir}^{\left( a\right) }$ and $X_{f}^{\left(a\right)}$ are related via the relation
\begin{equation}\label{Xrel}
 X_{\rm Casimir}^{\left(a\right)}(x_t)=(d-1) X_{f}^{\left(a\right)}(x_t) -\frac{1}{\nu} x_t \,\frac{d}{dx_t} X_{f}^{\left(a\right)}(x_t).
\end{equation}
The value of $X_{f}^{\left(a\right)}$ at the critical point is known as the Casimir amplitude $\Delta^{(a)}$, i.e. $\Delta^{(a)}\equiv X_{f}^{\left(a\right)}(x_t=0)$. On its turn, the excess free energy under antiperiodic conditions $f^{(a)}_{\rm ex}$ can be related to the one of the same system under periodic boundary conditions $f^{(p)}_{\rm ex}$ via the finite-size helicity modulus $\Upsilon(T,L)$ \cite{D93}
\begin{equation}\label{helmoddef}
f^{(a)}_{\rm ex}(T,L)=f^{(p)}_{\rm ex}(T,L)+\frac{\pi^2}{2L}\Upsilon(T,L),
\end{equation}
where $\Upsilon(T)\equiv \lim_{L\to\infty}\Upsilon(T,L)$ with $\Upsilon(T)\ge 0$. For the behavior of $\Upsilon(T,L)$ near $T_c$ the standard finite-size scaling theory states that
\begin{equation}\label{helmodfss}
\beta \Upsilon(T,L)  =L^{-(d-2)}X_\Upsilon\left(x_t\right),
\end{equation}
where $X_\Upsilon$ is universal scaling function. Actually, when $d=3$, a modification of Eq. (\ref{helmodfss}) has been suggested in \cite{P90} by Privman, who supposed the possibility of appearance of "resonant" logarithmic term due to the mutual influence of the regular and singular contributions in the helicity modulus
\begin{equation}\label{helmodfssPrivman}
\beta \Upsilon(T,L)  =L^{-1}\left[\tilde{X}_\Upsilon\left(x_t\right)+\omega \ln (L/a)\right] + \Phi(T) L^{-1}+\cdots,
\end{equation}
where $\omega$ is an universal amplitude, while $\Phi(T)$ is a regular at $T_c$ function and $a$ is some characteristic microscopic length scales (e.g., the distance between the molecules of the correlated fluid, or the lattice spacing). The validity of this hypothesis has been checked in \cite{D93} on the example of the exactly solvable mean-spherical model. No logarithmic corrections of the type predicted in (\ref{helmodfssPrivman}) have been found. Let us recall that in the case of superfluids
$(n =2, d= 3)$ the helicity modulus $\Upsilon$ is proportional \cite{FBJ73} to the superfluid
density fraction $\varrho$, namely $\varrho=(m/\hbar)^2 \Upsilon(T)$ with $m$ being the mass of the helium
atom, and is directly measurable (for experiments measuring $\varrho$  in thin films of
$^4$He see, e.g., Refs. \cite{RGB89} and \cite{GR91}). In fact, (\ref{helmodfssPrivman}) was  proposed in \cite{P90} as an attempt to improve the fit of the experimental data. It turns out, however, that the overall fit of the data is improved only in a very limited way,
provided one insists on the bulk value of $\nu$ in the scaling variable $x_t$. The scaling "data collapse" technique works well if one takes $\nu$  as an adjustable parameter not necessarily equal to the correlation length exponent. It also should be emphasized that one could expect additional complexity in the behavior of the finite-size scaling function of the
helicity modulus in the case of superfluid transitions in a film geometry;
nevertheless, the analysis of the experimental data shows no clear singularities or a jump in the finite-size scaling function \cite{RGB89,GR91}.

According to all the accumulated analytical and numerical evidences, see e.g. \cite{BDT2000}, \cite{privman} and references cited therein, when $x_t \gg 1$ both the excess free energy and the Casimir force under both periodic and antiperiodic boundary conditions in systems with short-ranged interactions is expected to tend to zero in an exponential-in-$L$ way. This is consistent with $\Upsilon(T)\equiv 0$ for $T\ge T_c$. When $x_t\to -\infty$ the same quantities tend to zero in an power-law-in-$L$ way.
This slow algebraic decay of $f_{\rm ex}$ (and of $F_{\rm Casimir}$) is, of
course, associated with the existence of soft modes in the system (spin waves) when $T<T_{c}$ and in the absence of an ordering external field destroying the $O(n)$ symmetry. This, in turn, will lead to a much greater (in comparison with the Ising-like case) Casimir (solvation) force when $T<T_c$ in $O(n)$ models. With respect to the Casimir force the last has not only been predicted theoretically, but has been also observed experimentally \cite{GC,GSGC2006} and, relatively recently, confirmed in a model study of the $XY$ model numerically  via Monte Carlo simulations \cite{H2007,VGMD2007}. The considered systems do not posses, however, a diffuse interface. When such am interface is present and $T<T_c$ from Eq. (\ref{helmoddef}) it is easy to see that
\begin{equation}
F_{\rm Casimir}^{\left(a\right) }(T< T_c, L \gg 1)\simeq  \frac{1}{2} \pi^{2} \Upsilon(T) L^{-2}. \label{fssclowTas}
\end{equation}
Since $\Upsilon(T)\ge 0$ the last implies that the force will be repulsive and much stronger, of the order of $L^{-2}$, than in systems with no diffuse interface where it is either of the order of $L^{-d}$, or smaller.

Since we consider film geometry, it is natural to allow for an anisotropy of the interactions in the system which reflects this geometry. To that aim we will take the interaction constant in the Hamiltonian along the surface, say $J_\parallel$, to be different from the one perpendicular to the film, say $J_\perp$. Since such anisotropy does not change the universality class of the bulk system one might naively expect that the scaling functions of the {\em finite} system $X_{\rm Casimir}^{\left(a\right)}$, $ X_{f}^{\left(a\right)}$  and $X_\Upsilon$ will be the same as for the isotropic system. Recently it has been argued, however, see Refs. \cite{CD2004,D2008}, that this is not true and that one shall expect these functions to be {\em nonuniversal} and depending on the ratio $J_\perp/J_\parallel$. It has been shown  \cite{CD2004,D2008} that the main reason for this state of affairs is the need of a generalization of the standard hyperuniversality hypothesis \cite{SFW72,A74,G75,HAHS76,W76,PAH91}. According to it, if $f_{b,\rm sing}(T)$ is the singular part of the bulk free energy density $f_b$ normalized per $k_B T$ and $\xi(T)$ is the bulk two-point correlation length in the isotropic system, then
\begin{equation}\label{hyperunstandard}
\lim_{T\to T_c^+} f_{b,\rm sing}(T) [\xi(T)]^d=Q,
\end{equation}
where $Q$ is a universal constant that characterizes the corresponding universality class. If now $f_{b,\rm sing}(T|J_\perp,J_\parallel)$  is the corresponding free energy in the anisotropic film system with $\xi_\parallel$ being the correlation length along the system surface and $\xi_\perp$ the one perpendicular to it, then the generalized hyperuniversality hypothesis states that
 \begin{equation}\label{hyperunhyp}
 \lim_{T\to T_c^+} f_{b,\rm sing}(T|J_\perp,J_\parallel) [\xi_\parallel(T)]^{d-1} \xi_\perp(T)=Q,
 \end{equation}
with $Q$ being the same universal quantity as in the isotropic case. Note that the new hypothesis involves two different correlation lengths, characterized by two different correlation length amplitudes, while the standard hypothesis deals with only one correlation length. Note also that the validity of (\ref{hyperunstandard}) is one of the main prerequisites for arguing the validity of the scaling hypothesis (\ref{fssfe}) by Privman and Fisher \cite{PF84}. It is, however, possible to relate the scaling functions of the anisotropic to that one of the isotropic system. Indeed, choosing the isotropic system to be such that its correlation length is equal to, say, $\xi_\perp$ from (\ref{hyperunstandard}) and (\ref{hyperunhyp}) one obtains that
\begin{equation}\label{relscaling}
 f_{b,\rm sing}(T|J_\perp,J_\parallel) \simeq \left[\frac{\xi_\perp(T)}{\xi_\parallel(T)}\right]^{d-1}
 f_{b,\rm sing}(T) , \quad T\to T_c^+,
 \end{equation}
and, thus one arrives at
\begin{equation}\label{relXf}
X_{f}^{\left(a\right)}(x_t|J_\perp,J_\parallel)=
\left[\frac{\xi_\perp(T)}{\xi_\parallel(T)}\right]^{d-1} X_{f}^{\left(a\right)}(x_t),
\end{equation}
where $\xi_\parallel$ and $\xi_\perp$ are the correlation lengths in the anisotropic system while $X_{f}^{\left(a\right)}(x_t)$ is the universal scaling function of the isotropic one. Of course, (\ref{relscaling}) and (\ref{relXf}) shall be considered only as plausible hypotheses which validity has to be verified. Note that, if valid, Eq. (\ref{relXf}) implies a relation of the Casimir amplitudes in the anisotropic and isotropic system
\begin{equation}\label{relDeltaGen}
\Delta_{\rm Casimir}(d|J_\perp,J_\parallel)=\left(\xi_\perp/\xi_\parallel\right)^{d-1} \Delta_{\rm Casimir}(d).
\end{equation}

In the current article on the example of the exactly solvable mean spherical model with $2<d<4$ we will demonstrate that in the anisotropic system with a diffuse interface  the scaling function $X_{\rm Casimir}^{\left(a\right)}$, $ X_{f}^{\left(a\right)}$  and $X_\Upsilon$ indeed depend, in addition on the scaling variable $x_t$, also on the ratio $J_\perp/J_\parallel$. This will lead, e.g., to nonuniversality of the Casimir amplitudes in such systems which are, however, simply related to the ones of the isotropic system via the relation (\ref{relDeltaGen}). We will determine the explicit form of the scaling function of the free energy, Casimir force and of the finite-size helicity modulus. For the case $d=3$ in the isotropic system we will find the universal values of these quantities at the critical point $T_c$ of the bulk system. We will also consider the case when the nearest neighbor interaction $J_\parallel$ along the film might be different from the one in orthogonal direction $J_\perp$.

The structure of the article is as follows. In Section \ref{SM} we define the model under consideration and provide some basic expressions needed for its treatment. The results for the finite size behavior of the free energy and of the Casimir force are presented in Section \ref{FECF}, where in subsection \ref{2d4} we present our general results for $2<d<4$, while in subsection \ref{d3} the explicit results for the important case of $d=3$ are given. Our findings about the behavior of the helicity modulus are contained in Section \ref{HM}. The article closes with a discussion and concluding remarks given in Section \ref{DCR}. Some technical details and results needed in the main text are derived in Appendixes \ref{asU} and \ref{powerrepr}.

\section{The Spherical Model}
\label{SM}

As stated above, we will study the finite-size behavior of an anisotropic system with a diffuse interface on the example of a spherical model embedded on a $d$-dimensional hypercubic
lattice ${\cal L} \in \mathbb{Z}^d$, where ${\cal L}=L_1\times L_2\times
\cdots L_d$. Let $L_i=N_i a_i, i=1,\cdots,d$, where $N_i$ is the
number of spins and $a_i$ is the lattice constant along the axis
$i$ with ${\bm e}_i$ being a unit vector along that axis, i.e. ${\bm e}_i.{\bm e}_j=\delta_{ij}$.
With each lattice site $\bm{r}$ one associates a real-valued spin
variable $S_{\bm r}$ which obeys the constraint
\begin{equation}
\label{constraint}
\frac{1}{\cal N} \sum_{{\bm r}\in {\cal L} } \langle S_{\bm r}^2
\rangle=1,
\end{equation}
where ${\cal N}=N_1 N_2 \cdots N_d$ is the total number of spins in the
system. The average in (\ref{constraint}) is with respect to the
Hamiltonian of the model
\begin{equation}
\label{Hamsm}
\beta{\cal H}=-\frac{1}{2}\beta\sum_{{\bm r}, {\bm
r}'}S_{\bm r}J({\bm r}, {\bm r}')S_{{\bm r}'}+s\sum_{{\bm
r}}S_{{\bm r}}^2.
\end{equation}
In the current article we will consider only the case of
nearest-neighbor interactions, i.e. we take $J({\bm r}, {\bm
r}')=J(|{\bm r}-{\bm r}'|)=\ J_i$, if ${\bm r}-{\bm r}'=\pm a_i {\bm
e}_i$, $i=1,\cdots,d$, and $J({\bm r}, {\bm r}')=0$ otherwise. Explicitly, one has $J({\bm r}, {\bm
r}')=\sum_{i=1}^d J_i[\delta({\bm r}-{\bm r}'-a_i {\bm
e}_i)+\delta({\bm r}-{\bm r}'+a_i{\bm
e}_i)]$.
Let periodic boundary conditions are applied across  directions ${\bm e}_i$, $i=1,\cdots(d-1)$, while antiperiodic boundary conditions, responsible for the creation of a diffuse interface within the system, are applied across ${\bm e}_d$. Generalizing for the considered here anisotropic case the results of \cite{BDT2000,SP85L,SP85,D98,D93} pertinent to an isotropic model, it can be shown that the free energy of the model (per unit spin) is given by \cite{rem1}
\begin{eqnarray}
\label{freeenergyspms}
&&\beta f^{(a)}(\beta,{\bm
N}|d,{\bm J})=-\frac{1}{2}\ln{\pi}\\
&& +\sup_{s>\hat{J}^{(a)}_{\rm max}}\left\{
 -s + \frac{1}{2 {\cal N}} \sum_{{\bm k}\in
{{\rm BZ}}_1} \ln{\left[ s- \frac{1}{2}\beta  \hat{J}^{(a)}({\bm
k}) \right]} \right\}, \nonumber
\end{eqnarray}
where ${\bm N}=(N_1,N_2, \cdots N_d)$, ${\bm J}=(J_1,J_2,\cdots, J_d)$,  $\hat{J}^{(a)}({\bm
k})$ is the Fourier transform of the interaction $\bm J$, i.e.
\begin{equation}
\hat{J}^{(a)}({\bm k})=\sum_{\bm r} J({\bm r})e^{i {\bm k}.{\bm r}},
\end{equation}
$\hat{J}^{(a)}_{\rm max}=\max_{{\bm k}} \hat{J}^{(a)}({\bm k})$, and the wave vector ${\bm k}=\{k_1, k_2, \cdots, k_d\}\in {{\rm BZ}}_1$ is with components  $k_i=2\pi n_i/L_i$, where
$n_i=0,\cdots,N_i-1$, $i=1,\cdots,(d-1)$, while $k_d=2\pi (n_d+1/2)/L_d$ with $n_d=0,\cdots,N_d-1$. Thus, explicitly one has
\begin{equation}
\hat{J}^{(a)}({\bm k})=2\sum_{i=1}^{d-1} J_i \cos\left(\frac{2 \pi n_i}{N_i}\right) + 2 J_d \cos\left(\frac{\pi (2 n_d+1)}{N_d}\right),
\end{equation}
and $\hat{J}^{(a)}_{\rm max}= 2\sum_{i=1}^{d-1}J_i+2J_d\cos(\pi/N_d)\equiv \hat{J}_0-2J_d[1-\cos(\pi/N_d)]$, with $\hat{J}_0=2\sum_{i=1}^{d}J_i$. Note that the ground state energy $\hat{J}^{(a)}_{\rm max}$ depends on $N_d$ and is twofold degenerate - it is reached for both $k_1=k_2=\cdots=k_{d-1}=k_d=0$ and $k_1=k_2=\cdots=k_{d-1}=0, k_d=N_d-1$.
The equation (\ref{constraint}) for the spherical
field $s$ reads
\begin{equation}
\frac{1}{2{\cal N}}\sum_{{\bm k}\in {\rm
BZ}_1}\frac{1}{s-\frac{1}{2}\beta \hat{J}^{(a)}({\bm k})}=1.
\label{sfe}
\end{equation}

We will be mainly interested in determination of the Casimir force and the helicity modulus within the considered model in a film geometry. For that aim let us take $J_1=J_2=\cdots=J_{d-1}=J_\parallel$, $N_1=N_2=\cdots=N_{d-1}=N_\parallel$, $J_d=J_\perp$, $N_d=N_\perp$ and to perform the limit $N_\parallel\to\infty$, i.e. to consider a system with a film geometry in which all the interactions in directions parallel to the film surface are equal (to $J_\parallel$) but possible different from the interaction in the direction perpendicular to the surface (which is $J_\perp$). Then,  Eqs. (\ref{freeenergyspms}), and  (\ref{sfe})
become
\begin{eqnarray}
\label{freeenergyspmfilm}
\beta f^{(a)}(\beta, N_\perp|d,{\bm J})&=&\frac{1}{2}\ln{\frac{K}{2\pi}}-\frac{1}{2}K \frac{\hat{J}^{(a)}_{\rm max}}{\hat{J}_0}\\
& & +\sup_{w>0}
\Big\{U^{(a)}(w,N_\perp|d,{\bm J})-\frac{1}{2}K w\Big\}, \nonumber
\end{eqnarray}
\begin{equation}
K=\frac{1}{N_\perp} \sum_{k_d}\int_{{\bm k}_\parallel \in
{\rm BZ}_1}^{(d-1)}\frac{1}{w+\omega^{(a)}({\bm k}_\parallel,k_d|d,{\bm J})},
\label{sfew}
\end{equation}
correspondingly, where ${\bm k}=({\bm k}_\|, k_d)$ with ${\bm k}_\|= (k_1,k_2, \cdots, k_{d-1})
$, $K=\beta \hat{J}_0$,
\begin{eqnarray}\label{defofU}
\lefteqn{U^{(a)}(w,N_\perp|d,{\bm J})}\nonumber\\
  & &=\frac{1}{2N_\perp} \sum_{k_d}\int_{{\bm k}_\parallel \in
{\rm BZ}_1}^{(d-1)} \ln{\left[ w+\omega^{(a)}({\bm k}_\parallel,k_d|d,{\bm J})\right]},\quad
\end{eqnarray}
with
\begin{equation}\label{defofomega}
\omega^{(a)}({\bm k}|d,{\bm J})=\left[\hat{J}^{(a)}_{\rm max}- \hat{J}^{(a)}({\bm k})\right]/\hat{J}_0\ge 0,
\end{equation}
\begin{equation}
\int_{{\bm k}\in
{\rm BZ}_1}^{(d-1)} \equiv \prod_{i=1}^{d-1}\int_{0}^{2\pi}\frac{dk_i}{2\pi},
\end{equation}
and we have replaced the spherical field $s$ by another field $w$, defined as
\begin{equation}\label{defofw}
w=2s/K-\hat{J}^{(a)}_{\rm max}/\hat{J}_0.
\end{equation}
Here
\begin{equation}\label{gsefs}
  \hat{J}^{(a)}_{\rm max}=2(d-1)J_\parallel+2J_\perp \cos (\pi/N_\perp)
\end{equation}
is the ground state energy of the finite system under antiperiodic boundary conditions, while
\begin{equation}\label{gseis}
   \hat{J}_0=2(d-1)J_\parallel+2J_\perp
\end{equation}
is the ground state energy of the infinite one and, thus,
\begin{equation}\label{Jinbs}
\hat{J}^{(a)}_{\rm max}/\hat{J}_0=(d-1)b_\parallel+b_\perp \cos(\pi/N_\perp),
\end{equation}
where
\begin{equation}\label{bperp}
b_\perp=J_\perp/\sum_{i=1}^d J_i
\end{equation}
and
\begin{equation}\label{bpar}
   b_\parallel=J_\parallel/\sum_{i=1}^d J_i
\end{equation}
reflect the asymmetry in the interaction.

Eqs. (\ref{freeenergyspmfilm}) -- (\ref{defofw}) provide the basis for the investigation of the behavior of the Casimir force within mean-spherical model in the presence of a diffusive interface in the system.

\section{Finite-size behavior of the free energy and the Casimir force}
\label{FECF}

\subsection{General results for the case $2<d<4$}
\label{2d4}

From Eq. (\ref{freeenergyspmfilm})  for the excess free energy $\beta f^{(a)}_{\rm ex}(\beta,N_\perp|d, {\bm b})=N_\perp[\beta f^{(a)}(\beta,N_\perp|d, {\bm b})-\beta f_b(\beta|d, {\bm b})]$ one obtains
\begin{eqnarray}\label{fexfilm}
\lefteqn{\beta f^{(a)}_{\rm ex}(\beta,N_\perp|d, {\bm b})=N_\perp \Bigg[\frac{1}{2}\, b_\perp K  \left(1-\cos \frac{\pi}{N_\perp}\right)}\\
&& -\frac{1}{2}K\left(w-w_b\right) +U^{(a)}\left(w,N_\perp|d, {\bm b}\right)-U_d\left(w_b| {\bm b}\right)\Bigg],\nonumber
\end{eqnarray}
where $f_b(\beta|d, {\bm b})\equiv \lim_{N_\perp\to\infty} f(\beta,N_\perp|d, {\bm b})$, ${\bm b}=(b_\|, \cdots, b_\|,b)$,  $w\equiv w(K,N_\perp|d, {\bm b})$ is the solution of Eq. (\ref{sfew}), and $w_b\equiv w_b(K|d, {\bm b})$ is the $\lim_{N_\perp\to\infty}$ limit of  $w_b(K|d, {\bm b})$, i.e. $w_b(K|d, {\bm b})= \lim_{N_\perp\to\infty} w(K,N_\perp|d, {\bm b})$. As it is well known, see e.g. \cite{BDT2000}, for $K<K_c=W_d(0|{\bm b})$ the spherical filed $w_b$ is solution of the equation
\begin{equation}\label{sfbulk}
    K=W_d(w_b|{\bm b}),
\end{equation}
where, for $w\ge 0$,
\begin{equation}\label{Wdef}
  W_d(w|{\bm b})=  \frac{1}{2}  \int_{{\bm k}\in
{\rm BZ}_1}^{(d)}\frac{1}{w+\omega({\bm k}|d, {\bm b})},
\end{equation}
and $w_b=0$, when $K\ge K_c$. In Eq. (\ref{fexfilm}) $U_d(w|{\bm b})=\lim_{N_\perp\to \infty}U^{(a)}(w,N_\perp|d, {\bm b})$ which limit, according to Eq. (\ref{defofU}), reads
\begin{equation}\label{Ubulk}
U_d(w|{\bm b})=\frac{1}{2}  \int_{{\bm k}\in
{\rm BZ}_1}^{(d)}\ln[w+\omega({\bm k}|d, {\bm b})].
\end{equation}
Note that it does not depend on the boundary conditions.
Obviously, the only nontrivial $N_\perp$ dependence in $f^{(a)}_{\rm ex}$ stems from the size dependence of the spherical field $w$ and from the asymptotic behavior of $U^{(a)}(w,N_\perp|d, {\bm b})$ on $N_\perp$ for $N_\perp \gg 1$. Let us now study these dependencies in detail.

Using the identity
\begin{equation}\label{idforU}
\ln a=\int_0^\infty \frac{dx}{x}\left(e^{-x}-e^{-a x}\right)
\end{equation}
one can rewrite Eq. (\ref{defofU}) into the form
\begin{eqnarray}\label{Ufactor}
U^{(a)}(w,N_\perp|d, {\bm b})&=&\frac{1}{2}\int_0^\infty \frac{dx}{x} \bigg\{e^{-x}-e^{-w x} S_{N_\perp}^{(a)}(x b_\perp)\times\nonumber \\ && \lefteqn{\times\left[e^{-x b_\parallel} I_{0}(x b_\parallel) \right]^{d-1}\bigg\}},
\end{eqnarray}
where
\begin{equation}\label{snadef}
S_N^{\rm (a)}(z)=\frac{1}{N} \sum_{n=0}^{N-1} \exp \left [ -z \left( \cos \frac{\pi}{N}-\cos \frac{\pi (2n+1}{N} \right) \right ].
\end{equation}
With the help of the identity
\begin{equation}\label{snasnp}
S_N^{\rm (a)}(z)=\exp \left[z \left(1-\cos \frac{\pi}{N} \right) \right] \left[2 S_{2N}^{\rm (p)}(z)-S_{N}^{\rm (p)}(z)\right],
\end{equation}
where
\begin{equation}\label{defofsp}
   S_{N}^{\rm (p)}(z) =\frac{1}{N} \sum_{{n=0}}^{N-1} \exp\left[-z\left( 1- \cos \frac{2\pi n}{N} \right) \right]
\end{equation}
the problem for determination of the asymptotic behavior of the sum $S_N^{\rm (a)}(x)$ when $N\gg 1$, which characterizes the {\it antiperiodic} boundary conditions, can be reduced to the determination of the asymptotic behavior of the sum $S_N^{\rm (p)}(x)$, which is pertinent to systems with {\it periodic} boundary conditions. It can be shown that \cite{D93}
\begin{widetext}
\begin{equation}\label{snaas}
S_N^{\rm (a)}(x)\simeq \left
\{\begin{minipage}[h]{12cm}
\begin{eqnarray*}
S_N^{\rm +(a)}&=& \frac{2}{N}+\frac{2}{N}R^{\rm (+)}\left(\frac{\pi^2}{2N^2}x\right)-v(x/2), \qquad x\ge N^2\\
S_N^{\rm -(a)}&=& \exp \left[x (1-\cos \frac{\pi}{N})\right]\left[e^{-x}I_0(x)+
\sqrt{\frac{2}{\pi x}}\;R^{\rm (-)}\left(\frac{2N^2}{x}\right)\right], \qquad  x\le N^2,
\end{eqnarray*} \end{minipage} \right.
\end{equation}
\end{widetext}
where
\begin{equation}\label{rap}
R^{\rm (+)}(x)=\sum_{n=1}^{\infty}e^{-4n(n+1)x},
\end{equation}
\begin{equation}\label{ram}
R^{\rm (-)}(x)=2\sum_{n=1}^{\infty}e^{-n^2 x}-\sum_{n=1}^{\infty}e^{-n^2 x/4},
\end{equation}
\begin{equation}\label{v}
v(x)=\frac{1}{\sqrt{4\pi x}}\left[1-{\rm erf} \left(\pi \sqrt{x}\,\right)\right].
\end{equation}
In addition, with the help of the Poisson identity, one can easily check that the following equivalent representations of functions $R^{\rm (+)}(x)$ and $R^{\rm (-)}(x)$ are valid
\begin{equation}\label{rapeq}
R^{\rm (+)}(x)= \frac{1}{2}e^x \theta_2 \left(0,e^{-4x}\right)-1,
\end{equation}
\begin{equation}\label{rameq}
R^{\rm (-)}(x)=\sum_{n=1}^{\infty}(-1)^n e^{-n^2 x/4}=\frac{1}{2}\left[\theta_4 \left(0,e^{-x/4}\right)-1\right],
\end{equation}
where $\theta_2(x)$ and $\theta_4(x)$ are the corresponding theta functions.

If one insists on using only the second asymptote in Eq. (\ref{snaas}) as the one valid for all $x$, see, e.g. Ref. \cite{SP85}, then the corresponding result for $U^{(a)}(w,N_\perp|d, {\bm b})$ reads
\begin{eqnarray}
\label{Uparas}
  U^{(a)}\left(w,N_\perp|d, {\bm b}\right)  &=& U_d\left(\tilde{w}|{\bm b}\right)  -\frac{N_\perp^{-d}}{(4\pi)^{d/2}} \left(\frac{b_\perp}{b_\parallel}\right)^{(d-1)/2}\times\nonumber \\&& \times\int_0^\infty \frac{dx}{x}x^{-d/2}e^{-\tilde{y}x} R^{\rm (-)}\left(\frac{1}{x}\right),\nonumber\\&&
\end{eqnarray}
where
\begin{equation}\label{wtildedef}
\tilde{w}=w-b_\perp\left(1-\cos\frac{\pi}{N_\perp} \right),
\end{equation}
\begin{equation}\label{ytildedef}
\tilde{y}=y-\pi^2, \qquad y=(2N_\perp^2/b_\perp)\, w
\end{equation}
and, see Eq. (\ref{Ufactor}),
\begin{eqnarray}\label{Ufactorbulk}
U_d(\tilde{w}|{\bm b})=\frac{1}{2}\int_0^\infty \frac{dx}{x} \bigg \{e^{-x}-e^{-\tilde{w} x} \left[e^{-x b_\perp} I_{0}(x b_\perp) \right]\times\nonumber\\\times\left[e^{-x b_\parallel} I_{0}(x b_\parallel) \right]^{d-1}\bigg \}.\nonumber\\
\end{eqnarray}
Using the representation (\ref{Ufactorbulk}) it can be shown \cite{DK2004}
that when $\tilde{w} \to 0^+$ one has
\begin{eqnarray}\label{Udas}
    U_d(\tilde{w}|\bm b)&=&U_d(0|{\bm b})+\frac{1}{2}\tilde{w}\; W_d(0|{\bm b})\nonumber\\&&-\frac{1}{2}\frac{\Gamma(-d/2)}{(2\pi)^{d/2}\prod_{i=1}^d\sqrt{b_i}} \,\tilde{w}^{d/2}+\cdots,\qquad
\end{eqnarray}
with the dots representing terms of higher order than those retained in the expression. From Eqs. (\ref{Uparas}), (\ref{Udas}) and with the help of the representation (\ref{rameq}), for the finite-size part $U\left(w,N_\perp|d, {\bm b}\right)$ of the free energy in the limit $\tilde{w}\to 0^+$ and, thus $\tilde{y}\ge 0$ - see Eq. (\ref{ytildedef}), one obtains
\begin{eqnarray}
\label{Ufinite_standard}
  \lefteqn{U^{(a)}\left(w,N_\perp|d, {\bm b}\right)}\nonumber\\&=& U_d(0|{\bm b})+ \frac{1}{4}b_\perp \tilde{y}\; W_d(0|{\bm b}) N_\perp^{-2}
   -N_\perp^{-d}\left(\frac{b_\perp}{b_\parallel}\right)^{(d-1)/2}\times\nonumber\\&& \times\tilde{y}^{d/2}
  \left\{ \frac{1}{2}\frac{\Gamma(-d/2)}{(4\pi)^{d/2}}  + \frac{2}{(2\pi)^{d/2}} \sum_{n=1}^{\infty}(-1)^n \frac{K_{d/2}(n\sqrt{\tilde{y}})}{(n\sqrt{\tilde{y}})^{d/2}}\right\}\nonumber\\&&+O(N_\perp^{-4}).
\end{eqnarray}
Then, from Eqs. (\ref{fexfilm}), (\ref{Udas})  and (\ref{Ufinite_standard}) for the excess free energy one derives the final result
\begin{eqnarray}
\label{fex_standard_final}
  \beta f^{(a)}_{\rm ex}(\beta,N_\perp|d, {\bm b}) &=& N_\perp^{-(d-1)}\left(\frac{b_\perp}{b_\parallel}\right)^{(d-1)/2}  \Bigg \{ \frac{1}{4} x_t(\tilde{y}-y_b)\nonumber\\&&-\frac{1}{2}\frac{\Gamma(-d/2)}{(4\pi)^{d/2}} \left(\tilde{y}^{d/2}-y_b^{d/2} \right) \nonumber \\
   &&-\frac{2\tilde{y}^{d/2}}{(2\pi)^{d/2}} \sum_{n=1}^{\infty}(-1)^n \frac{K_{d/2}(n\sqrt{\tilde{y}})}{(n\sqrt{\tilde{y}})^{d/2}} \Bigg \},\nonumber\\
\end{eqnarray}
where $x_t$ is the temperature dependents scaling variable
\begin{equation}\label{xtdef}
    x_t=b_\perp \left(\frac{b_\parallel}{b_\perp}\right)^{(d-1)/2}\left(K_c-K\right)\,
    N_\perp^{1/\nu}, \qquad \nu=1/(d-2),
\end{equation}
\begin{equation}\label{ybdef}
y_b=(2 N_\perp^2 /b_\perp)w_b,
\end{equation}
with $w_b$ being the solution of the bulk spherical field equation (\ref{sfbulk}).

Let us now see  what is the correct answer when the complete asymptotic behavior, as given in Eq. (\ref{snaas}), is used for the determination of the excess free energy.

Using the asymptotes given by Eq. (\ref{snaas}) one obtains, see Appendix \ref{asU}, that:\begin{widetext}
\begin{eqnarray}
\label{Ufullas}
  U^{(a)}\left(w,N_\perp|d, {\bm b}\right)&=& U_d(0|{\bm b})+ \frac{1}{4}\, b_\perp \left(y-\pi^2\right)\; W_d(0|{\bm b})\, N_\perp^{-2}   -
  \frac{1}{2} N_\perp^{-d}\left(\frac{b_\perp}{b_\parallel}\right)^{(d-1)/2} \frac{1}{(4\pi)^{d/2}} \Bigg \{ \Gamma(-d/2)y^{d/2}\nonumber \\
   & &+\pi^2 \Gamma (1-d/2)y^{d/2-1}  + 2\sqrt{4\pi}\int_0^\infty \frac{dx}{x}x^{-(d-1)/2}e^{-y x}\left[1+R^{\rm (+)}\left(\pi^2 x\right)-\frac{1}{2 \sqrt{4\pi x}}(1+\pi^2 x)\right]\Bigg \},\nonumber\\
\end{eqnarray}
where $y=(2N_\perp^2/b_\perp) w \ge 0$. Then from Eqs. (\ref{fexfilm}) and  (\ref{Udas})  for the excess free energy one obtains
\begin{eqnarray}
\label{fex_new_final}
\beta f^{(a)}_{\rm ex}(\beta,N_\perp|d, {\bm b})&=&N_\perp^{-(d-1)}\left(\frac{b_\perp}{b_\parallel}\right)^{(d-1)/2} \Bigg \{ \frac{1}{4} x_t\left(y-\pi^2-y_b\right) \nonumber \\&&-\frac{1}{2}\frac{1}{(4\pi)^{d/2}} \bigg[\Gamma\left(-d/2\right)\left(y^{d/2}-y_b^{d/2}\right)+\pi^2 \Gamma\left(1-d/2\right)y^{d/2-1}\bigg] + I(y,d)\Bigg \},
\end{eqnarray}
\end{widetext}
where
\begin{eqnarray}
\label{Idef}
\lefteqn{I(y,d)}\nonumber\\&\equiv&-\frac{1}{(4\pi)^{(d-1)/2}}\int_{0}^{\infty}\mathrm{d}x\, x^{-(d+1)/2}e^{-yx}\!\bigg[1+R^{(+)}(\pi^{2}x)\nonumber\\&&-\frac{1+\pi^{2}x}{2\sqrt{4\pi x}}\bigg].\end{eqnarray}
The expression (\ref{fex_new_final}) has to be compared with Eq. (\ref{fex_standard_final}) that follows when one uses as asymptote of $S_N^{\rm (a)}$, when $N\gg 1$,  only the asymptote $S_N^{\rm -(a)}$ from Eq. (\ref{snaas}) (see, e.g., Ref.  \cite{SP85}). As we see, (\ref{fex_new_final}) and (\ref{fex_standard_final}) differ from each other. However, using the identity
\begin{equation}\label{identity_Rp_Rm}
    1+R^{\rm (+)}\left(\pi^2 x\right)=\frac{e^{\pi^2 x}}{\sqrt{4\pi x}}
    \left[\frac{1}{2}+R^{\rm (-)}\left(\frac{1}{x}\right)\right]
\end{equation}
one can show that when $d<4$ and $y\ge \pi^2$
\begin{eqnarray}\label{int_r}
I(y,d)
&=& -\tilde{y}^{d/2} \frac{2}{(2\pi)^{d/2}} \sum_{n=1}^{\infty}(-1)^n \frac{K_{d/2}(n\sqrt{\tilde{y}})}{(n\sqrt{\tilde{y}})^{d/2}}\nonumber\\&&-
\frac{1}{2}\frac{\Gamma\left(-d/2\right)}{(4\pi)^{d/2}} \left[\left(\tilde{y}^{d/2}-y^{d/2}\right) +\pi^2 \frac{d}{2}\,y^{d/2-1}\right]\nonumber\\
\end{eqnarray}
and, thus, expression (\ref{fex_new_final}) is equivalent to (\ref{fex_standard_final}) for $y \ge \pi^2$. In the opposite case, when $y<\pi^2$, one can use (\ref{fex_new_final}) or, equivalently, the analytical continuation of (\ref{fex_standard_final}). Therefore, although in the derivation of (\ref{fex_standard_final}) the incomplete asymptotic behavior of  sums involved has been used, which makes this derivation mathematically wrong, and the expansion (\ref{Udas}) of the bulk quantities has been applied, which is valid  only for $y\ge \pi^2$, Eq. (\ref{fex_standard_final}) is still valid and can be used for all $y\ge 0$ since this equations is equivalent to (\ref{fex_new_final}) which is obtained when one follows the proper mathematical procedures.

When $|y-\pi^2|<4\pi^{2}$ one can provide a representation of the integral $I(y,d)$ in terms of power series which is very convenient for analysis of its behavior for small values of the argument $y$. The corresponding representation is derived in Appendix \ref{powerrepr}, and reads
\begin{eqnarray}
\label{Ips}
I(y,d)&=&\frac{y^{(d-2)/2}}{2(4\pi)^{d/2}}\!\left[\pi^{2}\Gamma\!
\left(1-d/2\right)+y\,\Gamma\!\left(-d/2\right)\right]\nonumber\\&&
-\pi^{(d-1)/2}\sum_{m=0}^{\infty}a_m^{(d)}(\pi^{2}-y)^{m},
\end{eqnarray}
where the coefficients $a_m^{(d)}$ are given by\begin{equation}\label{coeffamd}
a_m^{(d)}=\frac{(2^{1-d}-2^{-2m})
\Gamma\!\left(m+\frac{1-d}{2}\right)\zeta(2m+1-d)}{\pi^{2m}m!}.
\end{equation}

From Eqs. (\ref{def}), (\ref{fex_standard_final})  and (\ref{fex_new_final}) for the Casimir force one obtains the two equivalent representations:
\begin{eqnarray}
\label{fcas_standard_final}
  \lefteqn{\beta F^{(a)}_{\rm Casimir}(\beta,N_\perp|d, {\bm b})}\nonumber\\&=& N_\perp^{-d}\left(\frac{b_\perp}{b_\parallel}\right)^{(d-1)/2}  \Bigg \{ \frac{1}{4} x_t(\tilde{y}-y_b)-(d-1)\times\nonumber\\&&\times\Bigg[\frac{1}{2}\frac{\Gamma(-d/2)}{(4\pi)^{d/2}} \left(\tilde{y}^{d/2}-y_b^{d/2} \right) \nonumber
   +\tilde{y}^{d/2} \frac{2}{(2\pi)^{d/2}}\times\\&&\times \sum_{n=1}^{\infty}(-1)^n \frac{K_{d/2}(n\sqrt{\tilde{y}})}{(n\sqrt{\tilde{y}})^{d/2}}\Bigg] \Bigg \},
\end{eqnarray}
in the derivation of which we have used the identity
\begin{equation}\label{Kidentty}
\frac{\partial}{\partial y}\Big[y^\mu K_\mu (ay)\Big]=-a y^\mu K_{\mu-1}(a y),
\end{equation}
and
\begin{eqnarray}
\label{fcas_new_final}
  \lefteqn{\beta F^{(a)}_{\rm Casimir}(\beta,N_\perp|d, {\bm b})}\nonumber\\ &=& N_\perp^{-d}\left(\frac{b_\perp}{b_\parallel}\right)^{(d-1)/2} \Bigg \{ \frac{1}{4} x_t\left(y-\pi^2-y_b\right) -(d-1)\times\nonumber\\&&\times\Bigg \{ \frac{1}{2}\frac{1}{(4\pi)^{d/2}} \bigg[\Gamma\left(-d/2\right)\left(y^{d/2}-y_b^{d/2}\right)\nonumber \\&&+\pi^2 \Gamma\left(1-d/2\right)y^{d/2-1}\bigg] - I(y,d)\Bigg \}\Bigg \}.
\end{eqnarray}
In (\ref{fcas_standard_final}) and (\ref{fcas_new_final}) the variables $y$ (or $\tilde{y}$) and $y_b$ satisfy the spherical field equations (\ref{sfew}) and (\ref{sfbulk}), respectively. It can be easily shown that these two equations can be rewritten in a scaling form. In the geometry of a film and under antiperiodic boundary condition the equation for $\tilde{y}$ reads
\begin{eqnarray}\label{yfinal}
-\frac{1}{2} x_t&=&\frac{\Gamma(1-d/2)}{(4\pi)^{d/2}} \tilde{y}^{d/2-1}
   +\tilde{y}^{d/2-1} \frac{2}{(2\pi)^{d/2}}\nonumber\\&&\times \sum_{n=1}^{\infty}(-1)^n \frac{K_{d/2-1}(n\sqrt{\tilde{y}})}{(n\sqrt{\tilde{y}})^{d/2-1}} ,
\end{eqnarray}
which is equivalent to
\begin{eqnarray}\label{ynewfinal}
-\frac{1}{2} x_t&=&\frac{\Gamma(1-d/2)}{(4\pi)^{d/2}} y^{d/2-1} + \frac{\Gamma(2-d/2)}{2^{d}\pi^{d/2-2}} y^{d/2-2}
   \nonumber\\&&+2\frac{d}{dy}I(y,d),
\end{eqnarray}
while the corresponding equation for $y_b$ is
\begin{equation}\label{ybfinal}
-\frac{1}{2} x_t=\frac{\Gamma(1-d/2)}{(4\pi)^{d/2}} y_b^{d/2-1}.
\end{equation}

Eqs. (\ref{xtdef}), (\ref{fcas_standard_final}), (\ref{fcas_new_final}), (\ref{yfinal}), and (\ref{ybfinal}) demonstrate that the Casimir force in a system with anisotropic interaction can be written in the form
\begin{equation}\label{Casscaling}
\beta F^{(a)}_{\rm Casimir}(\beta,N_\perp|d, {\bm b}) =N_\perp^{-d} \left(\frac{b_\perp}{b_\parallel}\right)^{(d-1)/2} X_{\rm Casimir}(x_t),
\end{equation}
where $X_{\rm Casimir}$ is a universal scaling function, provided a suitable definition of the scaling variables, see Eq. (\ref{xtdef}), is used. Note that $x_t$ is of the form $x_t=a_t({\bm b})\; t L^{1/\nu}$ which means that all the effect of the anisotropy of the type considered can be incorporated in the factor $\left(b_\perp / b_\parallel \right)^{(d-1)/2}=\left(J_\perp / J_\parallel \right)^{(d-1)/2}$ in front of the scaling function on the r.h.s. of Eq. (\ref{Casscaling}) and in the nonuniversal factor $a_t$ that enters in the definition of the temperature scaling variable $x_t$. Note that with respect to the Casimir amplitudes Eq. (\ref{Casscaling}) leads to the following relation between the amplitudes in the anisotropic and isotropic systems
\begin{equation}\label{CasAmplRel}
\Delta_{\rm Casimir}(d| J_\perp,J_\parallel) = \left(\frac{J_\perp}{J_\parallel}\right)^{(d-1)/2} \Delta_{\rm Casimir}(d).
\end{equation}
Note also that, because of the universality, the value of the Casimir amplitude in the isotropic system does not depend on $J\equiv J_\perp=J_\parallel$. In order to achieve a conformity with the relation (\ref{relDeltaGen}) one needs only to determine the ratio $\xi_\perp/\xi_\parallel$ in the anisotropic system. In fact, this already has been done in \cite{DK2004} with the result that
\begin{equation}\label{relxi}
\frac{\xi_\perp}{\xi_\parallel}=\sqrt{\frac{J_\perp}{J_\parallel}}.
\end{equation}
Inserting (\ref{relxi}) into (\ref{CasAmplRel}) one, indeed, immediately obtains (\ref{relDeltaGen}).

\subsection{Results for the case $d=3$}
\label{d3}

\begin{figure}[htb]
\includegraphics[angle=0,width=1.0\columnwidth,clip]{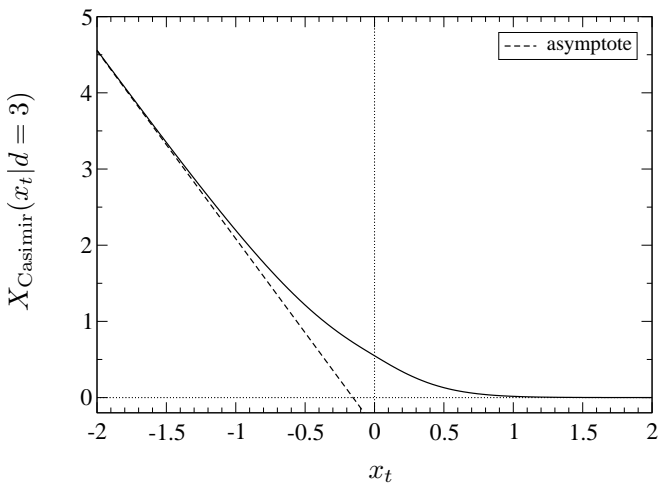}
\caption{The scaling function $X_{\rm Casimir}(x_t)$ of the Casimir force $F^{(a)}_{\rm Casimir}(\beta,N_\perp|d=3, {\bm b})$ for $d=3$. Note that $X_{\rm Casimir}(x_t)>0$ for all $x_t$. The asymptotic behavior of $X_{\rm Casimir}(x_t)$ for $x_t \ll -1$ is given according to Eq. (\ref{CasXas}). \label{Cas_plot}}
\end{figure}
Since $d=3$ is of special importance we will present some explicit results for this case. With $d=3$, the equations (\ref{fcas_standard_final}) and (\ref{fcas_new_final}) simplify to
\begin{eqnarray}
\label{fcas_standard_final_d3}
  \lefteqn{\beta F^{(a)}_{\rm Casimir}(\beta,N_\perp|d=3, {\bm b})}\nonumber\\ &=& N_\perp^{-3}\left(\frac{b_\perp}{b_\parallel}\right)
  \Bigg \{ \frac{1}{4} x_t(\tilde{y}-y_b)-\frac{1}{6 \pi }\left(\tilde{y}^{3/2}-y_b^{3/2}\right)\nonumber\\&&
  -\frac{\sqrt{\tilde{y}}}{\pi}\text{Li}_2\left(-e^{-\sqrt{\tilde{y}}}\right)-
  \frac{1}{\pi}\text{Li}_3\left(-e^{-\sqrt{\tilde{y}}}\right)\Bigg\},
\end{eqnarray}
and
\begin{eqnarray}
\label{fcas_new_final_d3}
 \lefteqn{\beta F^{(a)}_{\rm Casimir}(\beta,N_\perp|d=3, {\bm b})}\nonumber\\ &=& N_\perp^{-3}\left(\frac{b_\perp}{b_\parallel}\right)
   \Bigg \{ \frac{1}{4} x_t\left(y-\pi^2-y_b\right)  -\frac{1}{6 \pi }\left(y^{3/2}-y_b^{3/2}\right)\nonumber\\&& +\frac{\pi }{4} y^{1/2} +2\, I(y,3)
  \Bigg \},
\end{eqnarray}
while the equations (\ref{yfinal}) and (\ref{ynewfinal}) for $\tilde{y}$ and $y$ become
\begin{equation}\label{yfinald3}
x_t=\frac{1}{2\pi} \sqrt{\tilde{y}}
+\frac{1}{\pi} \ln\left [1+e^{-\sqrt{\tilde{y}}} \right ],
\end{equation}
and
\begin{eqnarray}\label{yfinalpd3}
x_t&=&\frac{1}{2\pi} \sqrt{y}-\frac{\pi}{4 \sqrt{y}} +  \frac{1}{\pi}\int_0^\infty \frac{dx}{x}\, e^{-yx} \bigg[1+R^{(+)}(\pi^{2}x)\nonumber\\&&-\frac{1+\pi^{2}x}{2\sqrt{4\pi x}}\bigg],
\end{eqnarray}
respectively. Eq. (\ref{yfinald3}) can be explicitly solved in the form
\begin{equation}\label{yfd3solsf}
\sqrt{\tilde{y}} =2\, \rm{arccosh}\left [ \frac{1}{2}\, e^{\pi x_t} \right].
\end{equation}
At $T=T_c$, i.e. when $x_t=0$, this solution simplifies to
\begin{equation}\label{yfd3solcrit}
\sqrt{\tilde{y}} = \pm i \frac{2\pi}{3}.
\end{equation}
As it is well known \cite{BDT2000}, the scaling form of the solution of Eq. (\ref{ybfinal}) for $y_b$ for the infinite system with $d=3$ is
\begin{equation}\label{ybfinald3}
\sqrt{y_b}=\left \{\begin{array}{cc}
                     2\pi x_t,  & x_t\ge 0\\
                     0, & x_t \le 0
                   \end{array}
 \right..
\end{equation}
At $T=T_c$ with $y_b=0$, according to Eq. (\ref{ybfinald3}), and $\tilde{y}$ from Eq. (\ref{yfd3solcrit}) one can from Eq. (\ref{fcas_standard_final_d3}) obtain the Casimir amplitude in the form\begin{equation}\label{CasAmAdditional}
\Delta_{\rm Casimir}= \left(\frac{J_\perp}{J_\parallel}\right)\left[\frac{1}{3}\rm{Im} \left(
   \rm{Li}_2\left(\sqrt[3]{-
   1}\right)\right)-\frac{\zeta (3)}{6 \pi }\right],
\end{equation}
which, using the relation $\mathrm{Im}(\mathrm{Li}_2(e^{i\theta}))=\mathrm{Cl}_2(\theta)$ between the polylogarithm and the Clausen function (see, e.g., \cite{CC2007})\begin{equation}
\mathrm{Cl}_2(\theta)=\sum_{k=1}^{\infty}\frac{\sin(k\theta)}{k^2},
\end{equation}
can be written as\begin{eqnarray}\label{CasAmfinal}
\Delta_{\rm Casimir}&=&\left(\frac{J_\perp}{J_\parallel}\right) \left[\frac{1}{3} {\rm Cl}_2\left (\frac{\pi}{3} \right)-\frac{\zeta
   (3)}{6 \pi }\right] \nonumber \\&\simeq& 0.274543 \left(\frac{J_\perp}{J_\parallel}\right).
\end{eqnarray}
One can also determine the full temperature dependence of the Casimir force. For that aim, in Fig. \ref{Cas_plot} we present the scaling function $X_{\rm Casimir}(x_t)$ of the Casimir force $F^{(a)}_{\rm Casimir}(\beta,N_\perp|d=3, {\bm b})$ as a function of the temperature scaling variable $x_t$. We observe that $X_{\rm Casimir}(x_t)>0$ for all $x_t$, i.e. the Casimir force under antiperiodic boundary conditions is always a repulsive force. Furthermore, from Eqs. (\ref{fcas_standard_final_d3}), (\ref{yfinald3}) and (\ref{ybfinald3}) it is easy to check that when $x_t\gg 1$, one has $y, y_b\gg 1$ which lead to the result that the scaling function $X_{\rm Casimir}(x_t)$ decays exponentially fast to zero, while for $x_t\ll -1$ one has $y\to 0^+$, $y_b=0$ and that
\begin{equation}\label{CasXas}
X_{\rm Casimir}(x_t) \mathop{\approx}_{x_t\to-\infty}-\pi^2 x/4-\zeta(3)/\pi.
\end{equation}
As we will see below, the last equation, together with Eqs. (\ref{Casscaling}) and (\ref{Yanisotr}) - see below, lead to the conclusion that when $T\ll T_c$ the behavior of the Casimir force in systems with a diffuse interface in indeed given by Eq. (\ref{fssclowTas}).

\section{Helicity modulus}
\label{HM}

\subsection{General results for the case $2<d<4$}
\label{HM2d4}

The concept of the helicity modulus was introduced by Fisher {\it et al.} \cite{FBJ73}.
Fundamentally, the helicity modulus is a measure of the response of the
system to a helical or "phase-twisting" field. Alternatively, for an isotropic
system with $n$-component order parameter, where $n \ge 2$, one can consider the
helicity modulus to be the analogy of the surface tension or interfacial
free energy between two phases in a system with a scalar, i.e. $n= 1$ order
parameter (e.g., an Ising model). In other words - the helicity modulus is a measure of the increase of the energy of the system due to the existence of a diffuse interface within it. When in an $O(n)$, $n \ge 2$ such an interface is created by the application of antiperiodic boundary conditions the helicity modulus can be defined, e.g., as suggested in \cite{D93}
\begin{eqnarray}\label{helmod}
\lefteqn{\Upsilon(\beta,N_\perp|d, {\bm b})}\nonumber\\&\equiv&\frac{2N_\perp}{\pi^2}\left[f^{(a)}_{\rm ex}(\beta,N_\perp|d, {\bm b})-f^{(p)}_{\rm ex}(\beta,N_\perp|d, {\bm b})\right]\quad
\end{eqnarray}
where $f^{(p)}_{\rm ex}(\beta,N_\perp|d, {\bm b})$ is the excess free energy of the system under periodic boundary conditions when no such a diffuse interface exists. Obviously, the helicity modulus of the infinite system then simply is $\Upsilon(\beta|d, {\bm b})\equiv\lim_{N_\perp\to\infty}\Upsilon(\beta,N_\perp|d, {\bm b})$. Within the isotropic spherical model the corresponding result for $\Upsilon(\beta|d)$ is known, see, e.g., \cite{D93}
\begin{equation}\label{helbulk}
\beta\Upsilon(T|d)=\frac{1}{2d}(K-K_c).
\end{equation}
The needed information for $f^{(p)}_{\rm ex}(\beta,N_\perp|d, {\bm b})$ is also available, see, e.g., \cite{DK2004}
\begin{eqnarray}
\label{fex_standard_final_per}
  \lefteqn{\beta f^{(p)}_{\rm ex}(\beta,N_\perp|d, {\bm b})}\nonumber\\ &=& N_\perp^{-(d-1)}\left(\frac{b_\perp}{b_\parallel}\right)^{(d-1)/2}  \Bigg \{ \frac{1}{4} x_t(y_p-y_b)-\frac{\Gamma(-d/2)}{2(4\pi)^{d/2}}\times\nonumber\\&& \times\left(y_p^{d/2}-y_b^{d/2} \right)
   -y_p^{d/2} \frac{2}{(2\pi)^{d/2}} \sum_{n=1}^{\infty} \frac{K_{d/2}(n\sqrt{y_p})}{(n\sqrt{y_p})^{d/2}} \Bigg \},\nonumber\\
\end{eqnarray}
where $y_p$ is the solution of the equation
\begin{eqnarray}\label{yfinal_per}
-\frac{1}{2} x_t&=&\frac{\Gamma(1-d/2)}{(4\pi)^{d/2}} y_p^{d/2-1}
   +y_p^{d/2-1} \frac{2}{(2\pi)^{d/2}}\times\nonumber\\&&\times \sum_{n=1}^{\infty} \frac{K_{d/2-1}(n\sqrt{y_p})}{(n\sqrt{y_p})^{d/2-1}}.
\end{eqnarray}
Using Eqs. (\ref{fex_standard_final}) and (\ref{fex_standard_final_per}), for the finite-size scaling behavior of the helicity modulus we obtain
\begin{equation}\label{hel_mod_scaling}
\beta \Upsilon(\beta,N_\perp|d, {\bm b}) =  N_\perp^{-(d-2)}\left(\frac{J_\perp}{J_\parallel}\right)^{(d-1)/2} X_{\Upsilon}(x_t),
\end{equation}
where the scaling function of the helicity modulus $\Upsilon$ is
\begin{eqnarray}
\label{hel_mod_scaling_2d4}
 X_{\Upsilon}(x_t)  &=& \frac{2}{\pi^2}
   \Bigg \{ \frac{1}{4} x_t(\tilde{y}-y_p)-\frac{1}{2}\frac{\Gamma(-d/2)}{(4\pi)^{d/2}} \left(\tilde{y}^{d/2}-y_p^{d/2} \right) \nonumber \\
   &&-\frac{2}{(2\pi)^{d/2}} \bigg[ \tilde{y}^{d/2}  \sum_{n=1}^{\infty} (-1)^n \frac{K_{d/2}(n\sqrt{\tilde{y}})}{(n\sqrt{\tilde{y}})^{d/2}}\nonumber\\&&-y_p^{d/2}  \sum_{n=1}^{\infty} \frac{K_{d/2}(n\sqrt{y_p})}{(n\sqrt{y_p})^{d/2}} \bigg] \Bigg \},
\end{eqnarray}
where $\tilde{y}$ is the solution of Eq. (\ref{yfinal}), $y_p$ is the solution of Eq. (\ref{yfinal_per}), and $x_t$ is defined in Eq. (\ref{xtdef}). Taking into account that when $T<T_c$ and $N_\perp \gg 1$ one has $y_p\to 0^+$ and $y\to 0^+$, from Eqs. (\ref{hel_mod_scaling}) and (\ref{hel_mod_scaling_2d4}) one derives, within the spherical model, the behavior of the "bulk" helicity modulus in an anisotropic system
\begin{equation}\label{Yanisotr}
\beta\Upsilon(T|d, {\bm b})=\frac{1}{2}b_\perp (K-K_c).
\end{equation}

\subsection{Results for the case $d=3$}
\label{HMd3}

Since $d=3$ is of special importance we, similar to what we have done for the Casimir force in systems with diffuse interface, will present in more details explicit results for the finite-size behavior of the helicity modulus in this case. When $d=3$ the equations (\ref{hel_mod_scaling_2d4})  and (\ref{yfinal_per}) simplify to
\begin{eqnarray}
\label{hel_mod_scaling_d3}
  \lefteqn{\beta \Upsilon(\beta,N_\perp|d=3, {\bm b})}\nonumber\\ &=& N_\perp^{-1}\left(\frac{b_\perp}{b_\parallel}\right) \frac{2}{\pi^2}
   \Bigg \{ \frac{1}{4} x_t(\tilde{y}-y_p)-\frac{1}{12 \pi } \left(\tilde{y}^{3/2}-y_p^{3/2} \right) \nonumber \\
   &&-\frac{1}{2\pi} \bigg[ \sqrt{\tilde{y}}\,
   \text{Li}_2\left(-e^{-\sqrt{\tilde{y}}}\right)-\sqrt{y_p} \,
   \text{Li}_2\left(e^{-\sqrt{y_p}}\right)\nonumber\\&&+\text{Li}_3\left(
   -e^{-\sqrt{\tilde{y}}}\right)-\text{Li}_3\left(e^{-\sqrt{y_p}}\right) \bigg]
   \Bigg \},
\end{eqnarray}
and
\begin{equation}\label{yfinal_per_d3}
x_t=\frac{1}{2\pi} \sqrt{y_p}
+\frac{1}{\pi} \ln\left [1-e^{-\sqrt{y_p}} \right ],
\end{equation}
respectively. The solution of Eq. (\ref{yfinal_per_d3}) for periodic boundary conditions is
\begin{equation}\label{yfd3solsf_per}
\sqrt{y_p} =2\, \rm{arcsinh}\left [ \frac{1}{2}\, e^{\pi x_t} \right],
\end{equation}
which has to be compared with the corresponding solution for the antiperiodic boundary conditions, see Eq. (\ref{yfd3solsf}).
\begin{figure}[htb]
\includegraphics[angle=0,width=1.0\columnwidth,clip]{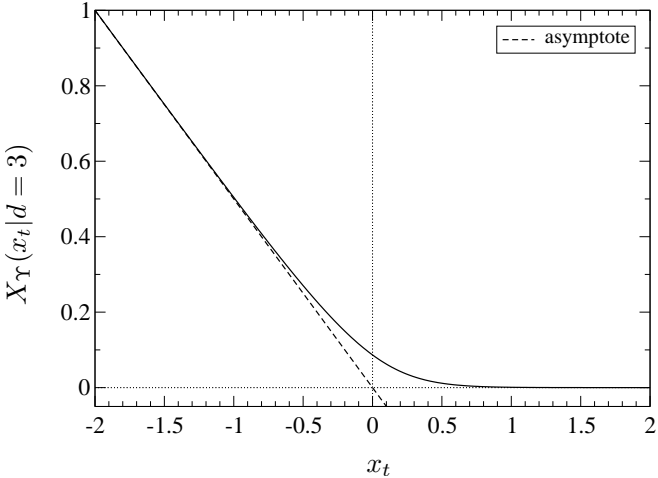}
\caption{The scaling function $X_{\Upsilon}(x_t)$ of the helicity modulus $\Upsilon(T,L)$ for $d=3$. One observes that it is a monotonically decreasing function of $x_t$. The asymptote of  $X_{\Upsilon}(x_t)$ for $x_t\ll -1$ is given in Eq. (\ref{helmodas}). \label{HMplot}}
\end{figure}

Let us determine the critical value of the finite-size helicity modulus $\Upsilon(\beta_c,N_\perp|d=3, {\bm b})$. Knowing the Casimir amplitude for antiperiodic boundary conditions $\Delta_{\rm Casimir}$ (see Eq. (\ref{CasAmfinal})) and that one under periodic boundary conditions \cite{D98} (see also \cite{DDG2006})
\begin{equation}\label{CasAmPer}
\Delta_{\rm Casimir}^{\rm per}=-\frac{2}{5\pi} \zeta(3)\simeq -0.153051,
\end{equation}
from  Eq. (\ref{helmod}) one obtains
\begin{eqnarray}\label{HMtc}
\lefteqn{\beta_c \Upsilon(\beta_c,N_\perp|d=3, {\bm b})}\nonumber\\&=&\frac{2}{\pi^2 N_\perp}\left(\frac{J_\perp}{J_\parallel}\right) \left[\Delta_{\rm Casimir}-\Delta_{\rm Casimir}^{\rm per}\right] \nonumber \\
&=& \frac{2}{\pi^2 N_\perp} \left(\frac{J_\perp}{J_\parallel}\right) \left[\frac{1}{3} {\rm Cl}_2\left(\frac{\pi}{3}\right)+\frac{7}{30\pi}\zeta(3)\right] \nonumber \\
&\simeq & 0.086649\; N_\perp^{-1} \left(\frac{J_\perp}{J_\parallel}\right).
\end{eqnarray}
Taking into account the relation $\varrho(T_c,L)=(m/\hbar)^2 \Upsilon(T,L)$ between the superfluid density fraction and the helicity modulus (strictly speaking this is valid only for $n=2$) one can obtain, within our model, an estimation of $\varrho(T_c,L)$ at $T_c$.

The dependence of the scaling function $X_\Upsilon(x_t)$ is shown in Fig. \ref{HMplot}. It is easy to show that $X_\Upsilon(x_t)$ decays exponentially fast for $x_t \gg 1$, while for $x_t\ll -1$ one derives that
\begin{equation}\label{helmodas}
X_\Upsilon(x_t) \mathop{\approx}_{x_t\to-\infty} -x_t/2.
\end{equation}
The asymptote of $X_\Upsilon$ for $T<T_c$ leads to Eq. (\ref{Yanisotr}) for the behavior of the helicity modulus within the anisotropic $O(n)$ models when $n\to\infty$ in the limit $\lim_{N_\perp\to \infty} \beta \Upsilon(\beta,N_\perp|d=3, {\bm b})$.

\section{Discussion and concluding remarks}
\label{DCR}
\begin{figure}[htb]
\includegraphics[angle=0,width=1.0\columnwidth,clip]{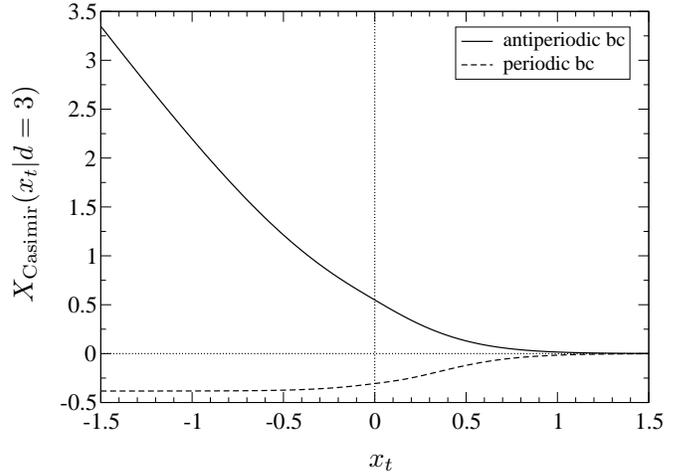}
\caption{The scaling functions $X_{\rm Casimir}(x_t)$ of the Casimir forces $F^{(a)}_{\rm Casimir}(\beta,N_\perp|d=3, {\bm b})$ and $F^{(p)}_{\rm Casimir}(\beta,N_\perp|d=3, {\bm b})$ for $d=3$. The difference is due to the contributions stemming from the helicity modulus. We see that this contribution is rather strong and dominates the behavior of the force under antiperiodic boundary conditions converting it from attractive (under periodic boundary condition) into a repulsive one (under antiperiodic boundary conditions). \label{Cas_plot_compare}}
\end{figure}
In the current article we studied the behavior of the Casimir force and the helicity modulus in anisotropic system with a diffuse interface as a function of the temperature. The interaction along the film is characterized via a coupling constant $J_\parallel$ while in the perpendicular to the film direction it is $J_\perp$. We have found that all scaling functions, including the Casimir amplitudes, depend on the ration $J_\perp/J_\parallel$ and are, thus, nonuniversal. More precisely, we have found that the Casimir force in a $d$-dimensional system, with $2<d<4$, can be written in the form, see Eq. (\ref{Casscaling})
\begin{eqnarray}\label{Casscalingfinal}
\lefteqn{F^{(a)}_{\rm Casimir}(T,N_\perp|d,J_\perp,J_\parallel)}\nonumber\\&=&(k_B T_c) \; N_\perp^{-d} \;  X_{\rm Casimir}(x_t|d,J_\perp,J_\parallel),
\end{eqnarray}
near the corresponding bulk critical temperature $T_c$, where $x_t$ is a properly defined temperature dependent scaling variable and the {\em nonuniversal} scaling function
 \begin{equation}\label{relaniso}
    X_{\rm Casimir}(x_t|d,J_\perp,J_\parallel)=\left(\frac{J_\perp}{J_\parallel}\right)^{(d-1)/2} X_{\rm Casimir}(x_t|d)
 \end{equation}
 can be related to $X_{\rm Casimir}(x_t|d)$, which is the {\em universal} scaling function characterizing the corresponding isotropic system. The explicit form of $ X_{\rm Casimir}(x_t|d)$, for $2<d<4$, is given in Eq. (\ref{fcas_standard_final}) and, equivalently, in (\ref{fcas_new_final}). Similar relations  can be written also for the helicity modulus, see Eq. (\ref{hel_mod_scaling})
\begin{equation}\label{hel_mod_scaling_final}
\Upsilon(T,N_\perp|d, J_\perp,J_\parallel) =  (k_B T_c) \; N_\perp^{-(d-2)} \; X_{\Upsilon}(x_t|d,J_\perp,J_\parallel),
\end{equation}
where, again, the {\em nonuniversal} scaling function $X_{\Upsilon}(x_t|d,J_\perp,J_\parallel)$
\begin{equation}\label{relhel}
X_{\Upsilon}(x_t|d,J_\perp,J_\parallel)=\left(\frac{J_\perp}{J_\parallel}\right)^{(d-1)/2} X_{\Upsilon}(x_t|d)
\end{equation}
can be related to  {\em universal} scaling function $X_{\Upsilon}(x_t|d)$ characterizing the corresponding isotropic system. The explicit form of $ X_{\Upsilon}(x_t|d)$, for $2<d<4$, is given in Eq. (\ref{hel_mod_scaling_2d4}).

From. Eq. (\ref{relaniso}) one obtains, see Eq. (\ref{CasAmplRel}), that
\begin{equation}\label{CasAmplRelFinal}
\Delta_{\rm Casimir}(d| J_\perp,J_\parallel) = \left(\frac{J_\perp}{J_\parallel}\right)^{(d-1)/2} \Delta_{\rm Casimir}(d).
\end{equation}
Since, within the spherical model, see Eq. (\ref{relxi}),
\begin{equation}\label{relxiFinal}
\frac{\xi_\perp}{\xi_\parallel}=\sqrt{\frac{J_\perp}{J_\parallel}}
\end{equation}
all the relations (\ref{relaniso}), (\ref{relhel}) and (\ref{CasAmplRelFinal}) are in full conformity with our general prediction given by Eqs. (\ref{relXf}) and (\ref{relDeltaGen}) which relate quantities of one anisotropic system to the corresponding ones in the isotropic system.

In addition to general expressions pertinent to the case $2<d<4$, we have also derived explicit results for the case $d=3$. The scaling function of the Casimir force is given in Eq. (\ref{fcas_standard_final_d3}) and, equivalently, in Eq. (\ref{fcas_new_final_d3}). The behavior of this function is visualized in Fig. \ref{Cas_plot}. The scaling function for the helicity modulus is presented in Eq. (\ref{hel_mod_scaling_d3}) and is depicted in Fig. \ref{HMplot}. For the value of the Casimir amplitude at $d=3$ one has, see Eq. (\ref{CasAmfinal}) \cite{C2008},
\begin{equation}\label{CasAmplFinalDisc}
\Delta_{\rm Casimir}= \left[ \frac{1}{3}{\rm Cl}_2\left (\frac{\pi}{3} \right)-\frac{\zeta
   (3)}{6 \pi }\right]\left(\frac{J_\perp}{J_\parallel}\right),
\end{equation}
while the value of the helicity modulus at $T_c$ is, see Eq. (\ref{HMtc}),
\begin{equation}
\label{HMtcFinal}
\beta_c\Upsilon(T_c,L)=\frac{2}{\pi^{2}} \left[\frac{1}{3}{\rm Cl}_2\Big(\frac{\pi}{3}\Big)+\frac{7\zeta(3)}{30\pi}\right]
\left(\frac{J_\perp}{J_\parallel}\right)L^{-1}.
\end{equation}

Let us note that both the Casimir amplitude, as well as the Casimir force are positive, i.e. they correspond to a repulsion between the plates of the system. Let us stress that this effect is solely due to the existence of a diffuse interface in the system. We recall that under periodic boundary conditions for $d=3$ and in the notations of the current article the Casimir force under periodic boundary conditions is given by the expression \cite{DK2004}
\begin{eqnarray}
\label{fcas_standard_per_d3}
  \lefteqn{\beta F^{(p)}_{\rm Casimir}(\beta,N_\perp|d=3, {\bm b})}\nonumber\\ &=& N_\perp^{-3}\left(\frac{b_\perp}{b_\parallel}\right)
  \Bigg \{ \frac{1}{4} x_t(y_p-y_b) -\frac{1}{6 \pi }\left(y_p^{3/2}-y_b^{3/2}\right)\nonumber\\&&
  -\frac{\sqrt{y_p}}{\pi}\text{Li}_2\left(e^{-\sqrt{y_p}}\right)-
  \frac{1}{\pi}\text{Li}_3\left(e^{-\sqrt{y_p}}\,\right)\Bigg\}.
\end{eqnarray}
The comparison between the force under antiperiodic and periodic boundary conditions is shown in Fig. \ref{Cas_plot_compare}. We observe that the contribution of the helicity energy is so strong that the Casimir force converts from being everywhere attractive (under periodic boundary conditions) into everywhere repulsive (under antiperiodic boundary conditions). This idea can eventually be used for practical purposes when applying some ordering external field might cause the border spins, dipoles, etc. to order in parallel or in antiparallel way to each other. Of course, by changing the degree of helicity the force will pass from being attractive through being zero into being repulsive. Obviously, it will be interesting to consider such a scenario in more details by say, studying a system under twisted at a given angle boundary conditions. We hope to return to this problem in a future work.

\acknowledgments

D.~D.\ would like to thank H.W.\ Diehl's group and Fachbereich Physik
of the Universit{\"a}t Duisburg-Essen for their hospitality at Campus
Duisburg.

We gratefully acknowledge the financial support of this work by the
Deutsche Forschungsgemeinschaft via Grant No.~Di-378/5, and the Bulgarian NSF
(Project F-1402).

\appendix

\section{Evaluation of $ U\left(w,N_\perp|d, {\bm b}\right)$}
\label{asU}
In the current appendix we prove the validity of Eq. (\ref{Ufullas}) for the behavior of $U\left(w,N_\perp|d, {\bm b}\right)$ when $N_\perp \gg 1$ and $0\le w\ll 1$. Because of the representation (\ref{Ufactor}) of $ U\left(w,N_\perp|d, {\bm b}\right)$ and the asymptotes (\ref{snaas}) of $S_N^{\rm (a)}(x)$ one divides the region of integration in two subregions - from $0$ to $a N_\perp^2$ and from $a N_\perp^2$ to infinity, where $a$ is a fixed real number such that $0<a<1$. Let us denote the integral over the first region (over "moderate" values of $x$) by $U_m$ and let $U_l$ is the integral over the "large" values of $x$, i.e. let
\begin{widetext}\begin{eqnarray}\label{Um}
U_m &\equiv& \frac{1}{2}  \int_0^{a N_\perp^2} \frac{dx}{x}\Bigg\{\exp(-x)-\exp\left\{-x \left[w- b_\perp \left(1-\cos \frac{\pi}{N_\perp}\right)\right]\right\}  \left[e^{-x b_\perp}I_0(x b_\perp)+ \sqrt{\frac{2}{\pi x b_\perp}}\;R^{\rm (-)}\left(\frac{2N_\perp^2}{x b_\perp}\right)\right]\times\nonumber\\&&\times\left[e^{-x b_\parallel} I_{0}(x b_\parallel) \right]^{d-1} \Bigg\}
\end{eqnarray}\end{widetext}
and
\begin{eqnarray}\label{Ul}
U_l &\equiv& \frac{1}{2}  \int_{a N_\perp^2}^\infty \frac{dx}{x}\Bigg\{e^{-x}-e^{-x w} \left[e^{-x b_\parallel} I_{0}(x b_\parallel) \right]^{d-1}\times\nonumber\\&&\times\left[ \frac{2}{N_\perp}+\frac{2}{N_\perp}R^{\rm (+)}\left(\frac{\pi^2 b_\perp}{2N_\perp^2}x\right)\right] \Bigg \}.
\end{eqnarray}
Obviously $U=U_l+U_m$. The evaluation of $U_l$ is straightforward. Since $x \gg 1$ in calculating $U_l$ one can use the large value asymptote of the Bessel function \cite{SP85}
\begin{equation}\label{sp}
    I_\nu(x)=\frac{\exp(x-\nu^2/2x)}{\sqrt{2\pi x}} \left[1+\frac{1}{8 x}+\frac{9-32\nu^2}{2! (8x)^2}+ \ \cdots\right]
\end{equation}
with the help of which one directly obtains that
\begin{eqnarray}\label{Ul_final}
U_l&=& -N_\perp^{-d}\left(\frac{b_\perp}{b_\parallel}\right)^{(d-1)/2}
\frac{1}{(4\pi)^{(d-1)/2}}\times\nonumber\\&& \times\int_{a b_\perp/2}^\infty \frac{dx}{x}x^{-(d-1)/2}e^{-y x}\left[1+R^{\rm (+)}\left(\pi^2 x\right) \right],\qquad
\end{eqnarray}
where $y$ is defined in Eq. (\ref{ytildedef}). Let us deal now with the term $U_m$. We divide this term into "bulk-like" contributions $U_{\rm m,b}$ and "finite-size" contributions $U_{\rm  m,fs}$ where $U_m=U_{\rm m,b}+U_{\rm  m,fs}$ with
\begin{eqnarray}\label{Umb}
U_{\rm m,b} &\equiv& \frac{1}{2}  \int_0^{a N_\perp^2} \frac{dx}{x}\Big\{e^{-x}-e^{-x \tilde{w}} \left[e^{-x b_\perp}I_0(x b_\perp)\right]\times\nonumber\\&&\times\left[e^{-x b_\parallel} I_{0}(x b_\parallel) \right]^{d-1} \Big\}
\end{eqnarray}
and
\begin{eqnarray}\label{Umfs}
U_{\rm m,fs} &\equiv& -   \int_0^{a N_\perp^2} \frac{dx}{x} e^{-x \tilde{w}} \frac{1}{\sqrt{2\pi x b_\perp}}\;R^{\rm (-)}\left(\frac{2N_\perp^2}{x b_\perp}\right)\times\nonumber\\&&\times\left[e^{-x b_\parallel} I_{0}(x b_\parallel) \right]^{d-1},
\end{eqnarray}
where
\begin{equation}\label{wtilde}
\tilde{w}=w- b_\perp \left(1-\cos \frac{\pi}{N_\perp}\right).
\end{equation}
It is straightforward to evaluate $U_{\rm m,fs}$. Due to the representation (\ref{rameq}), for all $x \ll N_\perp^2$ the corresponding contribution into the integral on the right-hand side of Eq. (\ref{Umfs}) will be exponentially small. Thus, one again can use in (\ref{Umfs}) the large-value asymptote (\ref{sp}) of the Bessel function $I_0(x)$ which leads to
\begin{eqnarray}\label{Umfsfinal}
U_{\rm m,fs}&=&-N_\perp^{-d}\left(\frac{b_\perp}{b_\parallel}\right)^{(d-1)/2}
\frac{1}{(4\pi)^{d/2}}\nonumber\times\\&& \times\int_0^{a b_\perp/2} \frac{dx}{x}x^{-d/2}e^{-\tilde{y} x}R^{\rm (-)}\left(\frac{1}{x}\right).
\end{eqnarray}
It remains now only to deal with the term $U_{\rm m,b}$. By subtracting  and adding, up to the linear in $x$ term,  the asymptote of $\exp\left[x b_\perp \left(1-\cos \frac{\pi}{N_\perp}\right)\right]$ for small values of $x$ one rewrites Eq. (\ref{Umb}) into the form
\begin{eqnarray}\label{Umbman}
U_{\rm m,b}\nonumber &=& \frac{1}{2}  \int_0^{a N_\perp^2} \frac{dx}{x}\Bigg\{e^{-x}-\left(1+\frac{1}{2}b_\perp \frac{\pi^2}{N_\perp^2}x\right)e^{-x w}\nonumber\\&&\times \left[e^{-x b_\perp}I_0(x b_\perp)\right]\left[e^{-x b_\parallel} I_{0}(x b_\parallel) \right]^{d-1} \Bigg\} \nonumber \\
& &  +\frac{1}{2}  \int_0^{a N_\perp^2} \frac{dx}{x}\left[\left(1+\frac{1}{2}b_\perp \frac{\pi^2}{N_\perp^2}x\right)e^{-x w}-e^{-x \tilde{w}}\right]\nonumber \\&&\times \left[e^{-x b_\perp}I_0(x b_\perp)\right]\left[e^{-x b_\parallel} I_{0}(x b_\parallel) \right]^{d-1}.
\end{eqnarray}
It is easy to understand that the integration over small values of $x$ in the second line of the above equation will provide contributions of the order of $O(N_\perp^{-4})$ which we will neglect, since we are only interested in contributions that are not smaller than $O(N_\perp^{-d})$, with $2<d<4$. Thus, in this integral one again can use the large value asymptote (\ref{sp}) of the Bessel function $I_0(x)$, which leads to\begin{widetext}
\begin{eqnarray}\label{Umbmanbulk}
U_{\rm m,b} &=& \frac{1}{2}  \int_0^{a N_\perp^2} \frac{dx}{x}\Bigg\{e^{-x}-e^{-x w} \left[e^{-x b_\perp}I_0(x b_\perp)\right]\left[e^{-x b_\parallel} I_{0}(x b_\parallel) \right]^{d-1} \Bigg\} -\frac{1}{4} b_\perp \frac{\pi^2}{N_\perp^2}  \int_0^{a N_\perp^2} dx  \; e^{-x w} \left[e^{-x b_\perp}I_0(x b_\perp)\right]\nonumber\times\\&&\times\left[e^{-x b_\parallel} I_{0}(x b_\parallel) \right]^{d-1}+  N_\perp^{-d}\left(\frac{b_\perp}{b_\parallel}\right)^{(d-1)/2}
\frac{1}{2} \frac{1}{(4\pi)^{d/2}} \int_0^{a b_\perp/2} \frac{dx}{x} \left[ e^{-x y}\left(1+\pi^2 x\right) - e^{-x \tilde{y}}\;\right] x^{-d/2}.
\end{eqnarray}
One can complete the integral in the first and second line of the above equation so that the integration is from $0$ to $\infty$ and to subtract the parts of integration from $a N_\perp^2$ to $\infty$. In the subtracted parts one can again use the large value asymptote (\ref{sp}) of the Bessel function $I_0(x)$. In this way one obtains
\begin{eqnarray}\label{Umbfinal}
\lefteqn{U_{\rm m,b}} &=& U_d(w|{\bm b})-\frac{1}{4}b_\perp \frac{\pi^2}{N_\perp^2} W_d(w|{\bm b})+N_\perp^{-d}\left(\frac{b_\perp}{b_\parallel}\right)^{(d-1)/2}
\frac{1}{2} \frac{1}{(4\pi)^{d/2}} \Bigg[\int_{a b_\perp/2}^\infty \frac{dx}{x} \;e^{-x y}\left(1+\pi^2 x\right) x^{-d/2} \nonumber \\
& &+  \int_0^{a b_\perp/2} \frac{dx}{x} \left[ e^{-x y}\left(1+\pi^2 x\right) - e^{-x \tilde{y}}\;\right] x^{-d/2}\Bigg].\nonumber\\
\end{eqnarray}
Expressing from Eq. (\ref{identity_Rp_Rm}) function $R^{\rm (-)}(x)$ in terms of $R^{\rm (+)}(x)$ and substituting the so-obtained representation in Eq. (\ref{Umfsfinal}) one obtains
\begin{eqnarray}\label{umsfinal_useful}
U_{\rm m,fs}&=&-N_\perp^{-d}\left(\frac{b_\perp}{b_\parallel}\right)^{(d-1)/2}
\frac{1}{(4\pi)^{(d-1)/2}} \int_0^{a b_\perp/2} \frac{dx}{x}x^{-(d-1)/2}e^{-y x}\left[1+R^{\rm (+)}\left(\pi^2 x\right)-\frac{1}{2}\frac{1}{\sqrt{4\pi x}}e^{\pi^{2}x}\right]\nonumber\\
&=&-N_\perp^{-d}\left(\frac{b_\perp}{b_\parallel}\right)^{(d-1)/2}
\frac{1}{(4\pi)^{(d-1)/2}} \Bigg \{ \int_0^{a b_\perp/2} \frac{dx}{x}x^{-(d-1)/2}e^{-y  x}\left[1+R^{\rm (+)}\left(\pi^2 x\right)-\frac{1}{2}\frac{1}{\sqrt{4\pi x}}\left(1+\pi^2 x\right)\right]\nonumber \\
&& + \frac{1}{2}\frac{1}{\sqrt{4\pi}} \int_0^{a b_\perp/2} \frac{dx}{x}x^{-d/2}e^{-y  x}\left[\left(1+\pi^2 x\right)- e^{\pi^{2}x}\right]\Bigg\}.
\end{eqnarray}
In a similar way, by adding and subtracting the  asymptote of $1+R^{\rm (+)}\left(\pi^2 x\right)$ for small values of the arument,  one can rewrite $U_l$ (see Eq. (\ref{Ul_final})) into the form
\begin{eqnarray}\label{Ul_final_helpful}
U_l&=& -N_\perp^{-d}\left(\frac{b_\perp}{b_\parallel}\right)^{(d-1)/2}
\frac{1}{(4\pi)^{(d-1)/2}} \Bigg\{ \int_{a b_\perp/2}^\infty \frac{dx}{x}x^{-(d-1)/2}e^{-y x}\left[1+R^{\rm (+)}\left(\pi^2 x\right)-\frac{1}{2}\frac{1}{\sqrt{4\pi x}}\left(1+\pi^2 x\right) \right]\nonumber \\
&& + \frac{1}{2}\frac{1}{\sqrt{4\pi}} \int_0^{a b_\perp/2} \frac{dx}{x}x^{-d/2}e^{-y  x}\left(1+\pi^2 x\right)\Bigg\}.
\end{eqnarray}\end{widetext}
By adding $U_{\rm m,b}$, $U_{\rm m,fs}$, and $U_l$ as given by Eqs. (\ref{Umbfinal}), (\ref{umsfinal_useful}) and (\ref{Ul_final_helpful}), respectively, one obtains, after using the representation (\ref{Udas}) for $U_d(w|\bm b)$, as well as the fact that $W_d(w|\bm b)=\partial U_d(w|\bm b)/\partial \omega $, the final result for  $U\left(w,N_\perp|d, {\bm b}\right)$ given in Eq. (\ref{Ufullas}) in the main text.

\section{Derivation of the series representation of $I(y,d)$}
\label{powerrepr}

In this appendix we derive the power series representation (\ref{Ips}) of the integral $I(y,d)$ defined in Eq. (\ref{Idef}).

First, let us note that using the representation (\ref{rap}) for the function $R^{\rm (+)}(x)$,
the integral $I(y,d)$ can be decomposed as $I(y,d)=I^{[1]}(y,d)+I^{[2]}(y,d)$,
where\begin{eqnarray}
I^{[1]}(y,d)&=&-\frac{1}{(4\pi)^{(d-1)/2}}\times\nonumber\\&&\times\sum_{n=1}^{\infty}\int_{0}^{\infty}dx\,x^{-(d+1)/2}e^{-yx}e^{-4\pi^{2}n(n+1)x}\quad\nonumber\\\end{eqnarray}
and\begin{eqnarray}
I^{[2]}(y,d)&=&-\frac{1}{(4\pi)^{(d-1)/2}}\times\nonumber\\&&\times\int_{0}^{\infty}dx\,x^{-(d+1)/2}e^{-yx}\!\left[1-\frac{1+\pi^{2}x}{2\sqrt{4\pi x}}\right].\quad\nonumber\\\end{eqnarray}
Employing dimensional regularization, the latter integral can be done
analytically and becomes
\begin{eqnarray}
I^{[2]}(y,d)&=&\frac{y^{(d-2)/2}}{2(4\pi)^{d/2}}\!\Big\{\pi^{2}\,\Gamma\left(1-d/2\right)
\\&&-4\sqrt{\pi y}\Gamma\left[(1-d)/2\right]+y\,\Gamma\left(-d/2\right)\Big\}.\nonumber
\end{eqnarray}
Introducing the variable $\tilde{y}\equiv y-\pi^{2}$, the integral
$I^{[1]}(y,d)$ can be written as\begin{eqnarray}
I^{[1]}(y,d)&=&-\frac{1}{(4\pi)^{(d-1)/2}}\times\nonumber\\&&\times\sum_{n=1}^{\infty}\int_{0}^{\infty}\frac{dx}{x^{(d+1)/2}}e^{-\tilde{y}x}e^{-[\pi^{2}+4\pi^{2}n(n+1)]x}\nonumber\\\end{eqnarray}
and upon replacing $\exp(-\tilde{y}x)$ by its Taylor series representation
the integral $I^{[1]}(y,d)$ becomes \begin{equation}\label{I1app}
I^{[1]}(y,d)=-\frac{1}{(4\pi)^{(d-1)/2}}\sum_{n=1}^{\infty}\sum_{m=0}^{\infty}b^{(d)}_{n,m}\frac{(-\tilde{y})^m}{m!}.\end{equation}
with the coefficients\begin{equation}
b^{(d)}_{n,m}(y)=\int_{0}^{\infty}dx\,x^{m-(d+1)/2}e^{-[\pi^{2}+4\pi^{2}n(n+1)]x}.
\end{equation}
Again in the sense of dimensional regularization the $x$-integration in the latter equation can be performed to give\begin{equation}
b^{(d)}_{n,m}(y)=\frac{\left(n+\frac{1}{2}\right)^{d-2m-1}\Gamma\!\left(m+\frac{1-d}{2}\right)}{(2\pi)^{2m+1-d}}.\end{equation}
Inserting this into Eq.~(\ref{I1app}), the $n$-summation can be done analytically leading to \begin{eqnarray}
I^{[1]}(y,d)&=&\pi^{(d-1)/2}\Bigg[2^{1-d}\sum_{m=0}^{\infty}\frac{(-\tilde{y})^{m}\Gamma\!\left(m+\frac{1-d}{2}\right)}{\pi^{2m}m!}\nonumber\\&&-\sum_{m=0}^{\infty}a^{(d)}_m(-\tilde{y})^{m}\Bigg]\end{eqnarray}with the coefficients $a^{(d)}_m$ defined in Eq.~(\ref{coeffamd}).
The first $m$-sum in square brackets can also
be done analytically and we obtain\begin{eqnarray}
I^{[1]}(y,d)&=&(4\pi)^{(1-d)/2}\Gamma\left[(1-d)/2\right](\tilde{y}+\pi^{2})^{(d-1)/2}\nonumber\\&&-\pi^{(d-1)/2}\sum_{m=0}^{\infty}a^{(d)}_m(-\tilde{y})^{m}.\end{eqnarray}
If we now add up $I^{[1]}(y,d)$ and $I^{[2]}(y,d)$ we arrive
at the power series representation (\ref{Ips}) of $I(y,d)$ given in the main text.
Note that no terms being nonanalytic with respect to $\tilde{y}$
are present, and furthermore that the radius of convergence of the
expansion is $|\tilde{y}|=|y-\pi^2|<4\pi^{2}$.


\begin{thebibliography}{88}

\bibitem{C48}H. B. G. Casimir, Proc. K. Ned. Akad. Wet. {\bf 51}, 793 (1948).

\bibitem{FdG78}M. E. Fisher and P. G. de Gennes,
C. R. Acad. Sc. Paris B {\bf 287}, 207 (1978).

\bibitem{evans}  R. Evans, in {\it Liquids at interfaces}, Les Houches
Session XLVIII, edited by J. Charvolin, J. Joanny and J. Zinn-Justin
(Elsevier, Amsterdam, 1990), p. 3.

\bibitem{K94} M. Krech, {\em The Casimir Effect in Critical Systems\/}
(World Scientific, Singapore, 1994).
\bibitem{BDT2000} J. G. Brankov, D. M. Danchev, and N. S. Tonchev,
{\em The Theory of Critical Phenomena in Finite-Size Systems -
Scaling and Quantum Effects\/} (World Scientific, Singapore, 2000).


\bibitem{KD92} M. Krech and S. Dietrich, Phys. Rev. Lett.
{\bf 66}, 345 (1991); Phys. Rev. A {\bf 46}, 1886 (1992); {\it
ibid} {\bf 46}, 1922 (1992), and references therein.

\bibitem{D98}  D.M. Danchev, Phys. Rev. E {\bf 53}, 2104 (1996);
{\it ibid} {\bf 58}, 1455 (1998).

\bibitem{M97} M. Krech, Phys. Rev. E {\bf 56}, 1642 (1997).

\bibitem{M99} M. Krech, J. Phys.: Condens. Matter {\bf 11}, R391
(1999).

\bibitem{GC} R. Garcia and M. H. W. Chan, Phys. Rev. Lett. {\bf 83},
1187 (1999);   Physica B {\bf 280}, 55 (2000);  J. Low Temp. Phys.
{\bf 121}, 495 (2000).

\bibitem{GCmixture} R. Garcia and M. H. W. Chan, Phys. Rev. Lett. {\bf 88}, 086101 (2002).

\bibitem{ML} A. Mukhopadhyay and B. M. Law, Phys. Rev. Lett. {\bf
83}, 772 (1999); Phys. Rev. E. {\bf 62}, 5201 (2000); {\it ibid}
{\bf 63}, 041605 (2001).

\bibitem{UBMCR03}  T. Ueno, S. Balibar, T. Mizusaki, F. Caupin, and
E. Rolley, Phys. Rev. Lett. {\bf 90}, 116102 (2003); R. Ishiguro and
S. Balibar, J. Low. Temp. Phys. {\bf 140}, 29 (2005).


\bibitem{GSGC2006} A. Ganshin, S. Scheidemantel, R. Garcia, and M. H.W.
Chan, Phys. Rev. Lett. {\bf 97}, 075301 (2006).




\bibitem{DKD2003} D. Dantchev, M. Krech, and S. Dietrich, Phys.
Rev. E {\bf 67}, 066120 (2003).

\bibitem{SHD03} {F. Schlesener, A. Hanke, and S. Dietrich,
J. Stat. Phys. {\bf 110}, 981 (2003).}

\bibitem{DK2004} D. Danchev and M. Krech, Phys. Rev. E {\bf 69}, 046119
(2004).



\bibitem{FYP05} M. Fukuto, Y. F. Yano, and P. S. Pershan, Phys. Rev.
Lett. {\bf 94}, 135702 (2005).

\bibitem{DDG2006} D. Dantchev, H.~W. Diehl, D. Gr{\"u}neberg, Phys.
Rev. E {\bf 73}, 016131 (2006).

\bibitem{DGS2006} H. W. Diehl, D. Gr\"uneberg, and M. A.
Shpot, Europhys. Lett. {\bf 75}, 241 (2006).

\bibitem{DSD2007} D. Dantchev, F. Schlesener and S. Dietrich, Phys. Rev. E {\bf 76}, 011121 (2007).

\bibitem{VGMD2007} O. Vasilyev, A. Gambassi, A. Maciolek, and S. Dietrich, Euro. Phys. Lett. {\bf 80}, 60009 (2007).

\bibitem{H2007} A. Hucht, Phys. Rev. Lett. {\bf 99}, 185301 (2007).

\bibitem{GD2008} D. Gr{\"u}neberg and H.~W. Diehl,  Phys.
Rev. B {\bf 77}, 115409 (2008).

\bibitem{HHGDB2008} C. Hertlein, L. Helden, A. Gambassi, S. Dietrich, and C. Bechinger, Nature {\bf 451}, 172 (2008).












\bibitem{privman}  V. Privman, in {\it Finite Size Scaling and Numerical
Simulations of Statistical Systems}, edited by V. Privman (World Scientific,
Singapore, 1990).

\bibitem{D93} D. Danchev, J. Stat. Phys. {\bf 73}, 267 (1993).

\bibitem{P90} V. Privman, J. Phys. A {\bf 23}, L711 (1990).






\bibitem{FBJ73} M. E. Fisher, M. N. Barber, and D. Jasnow, Phys. Rev. A {\bf 8}, 111 (1973).





\bibitem{RGB89} I. Rhee, F. M. Gasparini, and D. J. Bishop, Phys. Rev. Lett. {\bf 63}, 410 (1989).
\bibitem{GR91} F. M. Gasparini and I. Rhee, in {\it Progress in Low Temperature Physics}, Vol. 6, D. F. Brewer, ed. (North-Holland, Amsterdam, 1991), Chapter I.

\bibitem{CD2004} X. S. Chen and V. Dohm, Phys. Rev. {\bf 70}, 056136 (2004).

\bibitem{D2008} V. Dohm, cond-mat arXiv:0801.4096v2.

\bibitem{SFW72} D. Stauffer, M. Ferer, and M. Wortis, Phys. Rev. Lett. {\bf 29}, 345
(1972).

\bibitem{A74} A. Aharony, Phys. Rev. B {\bf 9}, 2107 (1974).

\bibitem{G75} P.R. Gerber, J. Phys. A {\bf 8}, 67 (1975).

\bibitem{HAHS76} P.C. Hohenberg, A. Aharony, B.I. Halperin, and E.D. Siggia, Phys. Rev. B {\bf 13}, 2986 (1976).

\bibitem{W76} F. Wegner, in {\it Phase Transitions and Critical Phenomena}, edited by
C. Domb and M.S. Green (Academic, New York, 1976), Vol.6, p.7.


\bibitem{PAH91} V. Privman, A. Aharony, and P.C. Hohenberg, in {\it Phase Transitions
and Critical Phenomena}, edited by C. Domb and J. L. Lebowitz (Academic, New York, 1991), Vol. 14, p. 1.

\bibitem{PF84} V. Privman  and M. E. Fisher,
Phys. Rev. B {\bf 30}, 322 (1984).





\bibitem{SP85L} S. Singh and R. K. Pathria, Phys. Rev. Lett. {\bf 55}, 347
(1985).

\bibitem{SP85} S. Singh and R. K. Pathria, Phys. Rev. B {\bf 85}, 4618
(1985).



\bibitem{rem1} The results presented in Eqs. (\ref{freeenergyspms}) and  (\ref{sfe})
can also be considered as a generalization toward antiperiodic boundary conditions of the correspondingg results of \cite{DK2004} for the spherical model with anisotropic interaction and periodic boundary conditions.



\bibitem{CC2007} J. Choi and D. Cvijovic, J. Phys. A {\bf 40}, 15019 (2007).

\bibitem{C2008} The numerical value of $\Delta_{\rm Casimir}\simeq 0.274543$ in the isotropic case has recently been also reported  in H. Chamati, arXiv:0805.0715.


\end{thebibliography}
\end{document}